\documentclass[%
reprint,
superscriptaddress,
 amsmath,amssymb,
 aps,
 prd,
]{revtex4-2}

\usepackage{graphicx}% Include figure files
\usepackage{dcolumn}% Align table columns on decimal point
\usepackage{bm}% bold math
\usepackage{color}

\usepackage{natbib}
\usepackage[colorlinks=true, urlcolor=blue, linkcolor=blue, citecolor=blue]{hyperref}

\begin{document}

\preprint{APS/123-QED}

%\title{The energy-energy correlators At low $ P_{T} $ \\ without reconstructing jets}% Force line breaks with \\
\title{Studying Energy–Energy Correlators in pp Collisions at the LHC with a Jet-Free Event-Topology Method}

\author{Yazhen Lin}\email{linyazz@mails.ccnu.edu.cn}
\affiliation{Key Laboratory of Quark and Lepton Physics (MOE) and Institute
	of Particle Physics, Central China Normal University, Wuhan 430079, China}

\author{Liang Zheng}\email{zhengliang@cug.edu.cn}
\affiliation{School of Mathematics and Physics, China University of
	Geosciences (Wuhan), Wuhan 430074, China}
\affiliation{Shanghai Research Center for Theoretical Nuclear Physics, NSFC and Fudan University, Shanghai 200438, China}
\affiliation{Key Laboratory of Quark and Lepton Physics (MOE) and Institute
	of Particle Physics, Central China Normal University, Wuhan 430079, China}	

\author{Zhongbao Yin}\email{zbyin@main.ccnu.edu.cn}
\affiliation{Key Laboratory of Quark and Lepton Physics (MOE) and Institute
	of Particle Physics, Central China Normal University, Wuhan 430079, China}


\begin{abstract}
%The energy-energy correlators (EEC) is a powerful tool typically applied within reconstructed jets to investigate the mechanisms of hadronization (the non-perturbative process where quarks and gluons form observable hadrons). However, in this study, we introduce a simple jet-free approach for EEC calculation. We compute the EEC directly by dividing the final-state charged particles into three azimuthal regions:Toward, Transverse, and Away, employing an region-based background subtraction method. This approach is crucial as it effectively obtains low $P_{T}$ EEC. This paper first compares this method with traditional jet reconstruction method, obtaining favorable results. It then utilizes the characteristics of leading particles to further analyze the dependence of the EEC on the trigger particle $P_{T}$, parton flavor, heavy flavor, and multiplicity.
 
 We present a jet-free approach for measuring energy–energy correlators (EEC) in proton–proton (pp) collisions at the Large Hadron Collider (LHC), employing an event-topology method that does not rely on explicit jet reconstruction. Using the leading charged hadron as a reference axis, the azimuthal plane is divided into Toward and Transverse regions, enabling a robust background subtraction and extending EEC measurements into the low $p_T$ regime where conventional jet-based approaches become unreliable. The method is validated through comparisons with conventional jet reconstruction results. We systematically explore the dependence of the EEC on the leading-particle transverse momentum and parton flavor. The observed scaling between the EEC peak position and the hard scale suggests that this topology-based EEC  captures effectively the transition between perturbative and non-perturbative QCD regimes. Distinct differences are found between quark- and gluon-initiated events, reflecting their different color charges and radiation patterns. Extending the analysis to heavy flavor, EECs triggered by leading charm mesons exhibit a suppressed magnitude and a peak shifted toward larger angular separations relative to inclusive charged-particle triggers, providing a direct manifestation of the dead-cone effect.
 This jet-free EEC framework offers a simple and experimentally robust tool for studying the scale and flavor dependence of the QCD dynamics, with promising applications to proton–nucleus and heavy-ion collisions at the LHC.
 
%\begin{description}
%\end{description}
\end{abstract}

\maketitle

%\tableofcontents

%\section{\label{sec:level1}First-level heading:\protect\\ The line
%break was forced \lowercase{via} \textbackslash\textbackslash}

\section{Introduction}

The energy-energy correlator (EEC) has emerged as a powerful and versatile observable in the study of quantum chromodynamics (QCD) and jet substructure physics~\cite{Hofman:2008ar,Larkoski:2017jix,Chen:2020vvp,Moult:2025nhu,Ali:1982ub}. By measuring the angular correlations of energy flow among final-state particles, the EEC provides a detailed probe of parton shower evolution and the complex interplay between perturbative and non-perturbative regimes~\cite{Gao:2019ojf,STAR:2025jut,Komiske:2022enw,ALICE:2024dfl}. A primary advantage of the EEC is its infrared and collinear (IRC) safety, which is inherently guaranteed by energy weighting. This ensures that the observable remains robust against soft radiation and collinear splittings, preserving a clean connection to the underlying partonic dynamics. Moreover, the EEC exhibits a well-defined angular scale dependence, allowing it to act as a dynamic resolution probe that maps QCD evolution across multiple energy scales~\cite{Lee:2022uwt,Hatta:2008tx,Brown:1981jv,PLUTO:1985yzc}.

At small angular separations, the EEC distribution is dominated by non-perturbative hadronization effects, which appear as a characteristic smearing of energy in space. The transition region between these small-angle correlations and the large-angle perturbative regime corresponds directly to the energy scale indicating the confinement transition from partons to hadrons. As the angular distance approaches or exceeds the jet radius, the distribution decreases rapidly due to geometric constraints, effectively marking the boundary of the correlated energy flow within the jet~\cite{CMS:2015wcf,Andres:2024hdd}. By comparing experimental measurements with high-precision theoretical calculations, the EEC allows for a quantitative determination of the hadronization corrections and offers unique insights into the mechanisms of color confinement. 

Beyond its role in jet substructure studies in proton–proton collisions, the EEC has broad applicability across different QCD environments. In heavy-ion collisions, the formation of a quark–gluon plasma (QGP) modifies the angular distribution of energy through medium-induced radiation and parton–medium interactions. The EEC is particularly sensitive to these effects, including large-angle correlations associated with the medium response and wake formation~\cite{Yang:2023dwc,CMS:2025ydi,Andres:2022ovj}. In jets initiated by heavy quarks, the finite quark mass suppresses small-angle gluon radiation through the dead-cone effect, leading to characteristic modifications of the EEC that directly encode fundamental mass-dependent QCD dynamics~\cite{Barata:2025uxp,Apolinario:2025vtx,Xing:2024yrb}. Furthermore, measurements in proton–nucleus and electron-nucleus collisions, allow for the isolation of cold nuclear matter effects, such as initial-state nuclear parton distribution function modifications, thereby providing a comprehensive picture of QCD evolution in diverse nuclear environments~\cite{Li:2021txc,Fu:2025gxu,Gonzalez:2022ulr,Barata:2024wsu,Electron-PositronAlliance:2025fhk}.

%State of the Art
%●Briefly discuss how EECs are traditionally measured within reconstructed jets, particularly at high transverse momentum (pt).
At collider experiments such as the Large Hadron Collider (LHC), the EEC measurements are usually performed within reconstructed jets, typically identified using IRC-safe clustering algorithms such as anti-$k_{t}$ algorithm~\cite{Cacciari:2008gp}. The standard approach involves summing over all pairs of particles within a reconstructed jet to obtain the EEC distribution.
High $p_{T}$ jet EEC distributions offer an ideal testing ground for perturbative QCD calculations, where background contributions are minimal and theoretical uncertainties are well controlled. However, at lower transverse momentum, traditional jet-based reconstruction becomes increasingly fraught with ambiguity. The presence of a dominant combinatorial background, arising from soft particle clustering and stochastic underlying-event fluctuations from multiple parton interactions (MPI), can either mimic or entirely mask the signals of genuine hard-scattered jets. Consequently, the standard jet-based EEC framework remains largely insensitive to the low $p_T$ regime. 
%This experimental barrier significantly impedes the access to critical information regarding jet energy loss and the non-perturbative dynamics of hadronization, which are most prominent at these lower $p_T$ scales.
To overcome these limitations, we propose a jet-free method that measures the EEC via event topology, eliminating explicit jet reconstruction. In this approach, the analysis uses the highest $p_T$ particle in the event as a reference to define the Toward and Transverse azimuthal regions~\cite{ATLAS:2019ocl,ALICE:2011ac,Peng:2025mpf,ALICE:2019mmy,Bencedi:2021tst}. This topological structure provides a natural axis for the event and enables a region-based background subtraction, allowing for a clean extraction of the low $p_T$ EEC signal~\cite{CDF:2004jod,Verma:2024fry}. The method retains the essential features of conventional EEC measurements while extending their applicability into the low $p_T$ domain.

%Outline
%Briefly state the structure of the paper
In this work, we apply this jet-free EEC framework to pp collisions at $\sqrt{s}=13$ TeV simulated with PYTHIA8, and validate it through comparisons with the standard jet-reconstruction-based method. 
%We then explore the sensitivity of the EEC to various physical factors, including the leading-particle $p_T$, parton flavor, heavy-flavor content. 
Through these studies, we demonstrate that the event topology based hadron method approach provides equivalent access to the perturbative/non-perturbative transition and parton shower flavor dynamics as the traditional jet based EEC observable. Crucially, this method offers a straightforward experimental implementation and can be extended to heavy-ion collisions, where it may serve as a more experimentally feasible probe sensitive to medium-induced modifications and quark–gluon plasma effects. The paper is organized as follows: Section.~\ref{sec:method} describes the event-topology method and background subtraction procedure in detail. Section.~\ref{sec:results} presents the main results and discusses the scaling feature and flavor sensitivity of this hadron based EEC observable. Section.~\ref{sec:summary} summarizes the conclusions and outlines possible extensions to different collision systems.

%Background substraction and matching method
\section{Methodology}\label{sec:method}
%Traditional EEC measurements are performed within reconstructed jets, where final state particles are clustered using various grouping algorithms. This method is believed to work well in high $p_T$ region, where jet boundaries are clearly defined and background contamination is minimal. However, the low $p_T$ region becomes poorly accessible in standard jet based EEC analysis since soft particle fluctuations and underlying event activity might obscure the genuine jet structure.

\begin{figure}[hbt!]
	\includegraphics[width=0.7\linewidth]{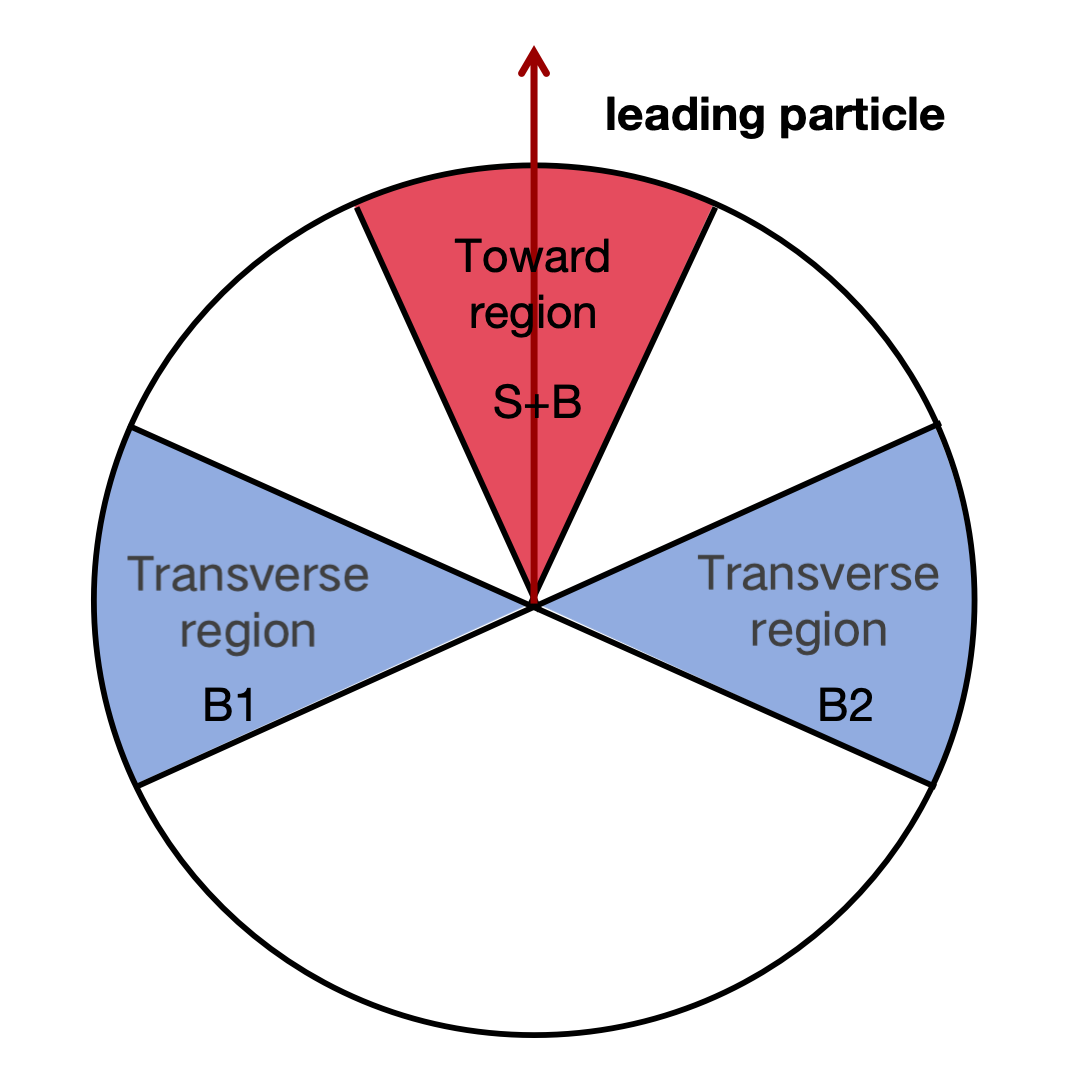}
	\caption{\label{fig:different_cone} The azimuthal plane is divided into different cones ($\eta$ is the same). The Toward cone (red) is defined around the leading particle. Two Transverse cones (blue) are made perpendicular to the leading particle.}
		%This diagram is merely illustrative and doesn't represent the actual cone size. The cone size are $R^{cone}$ = 0.5, with the maximum $\Delta\phi$ coverage approximately 28.65 degrees.
	
\end{figure}

The core of the hadron based method is to leverage the characteristic event topology of collisions to separate the hard scattering component from the softer contributions associated with the underlying event. The leading particle with the highest transverse momentum can be defined as a natural axis of the event. For each event with the leading particle satisfying the selection criteria $p_T$ greater than $p_T^{trig}$, we analyze the charged tracks around the leading particle within the angular region of a cone $\sqrt{(\phi-\phi_{leading})^2 + (\eta-\eta_{leading})^2}<R^{cone}$, which can be defined as the Toward region cone. The EEC distribution can be calculated as
\begin{equation}\label{eqn:theory_EEC}
\frac{d\Sigma_{EEC}}{dR_L}=\frac{1}{\sigma_{trig}}\sum_{i,j}\int \frac{p_{T,i}p_{T,j}}{p_{T,tot}^2}\delta(R_{L}-R_{L,ij})d\sigma_{ij},
\end{equation}
where $i,j$ run over all distinct pairs of particles in the set comprising the leading particle and all charged particles within the cone of radius $R^{cone}$ that satisfy $p_T>1$ GeV$/c$. The separation for any two particles in this set is defined by $R_{L,ij} = \sqrt{\Delta \phi_{ij}^2 + \Delta \eta_{ij}^2}$. The total transverse momentum $p_{T,tot}$ used for normalization can be obtained with the scalar sum of $p_T$ for each particle involved in the pair combinations within the Toward cone. To construct the correlation function, the pairwise product of transverse momenta $p_{T,i}p_{T,j}$, is employed as the weight for the energy flow. Utilizing $p_{T,tot}$ as a proxy of the total jet energy, this observable captures essentially the same underlying physics encoded in the jet based EEC measurements. Instead of employing explicit jet reconstructions, this approach maps the energy-weighted angular distribution of the parton shower by focusing on the correlated energy flow within the leading-particle neighborhood region. Following the definition in Eq.~\ref{eqn:theory_EEC}, the EEC distribution in experiment can thus be investigated varying with $R_L$ via
\begin{equation}\label{eqn:exp_EEC}
\begin{split}
\frac{d \Sigma_{EEC}(R_L)}{d R_L} &=\frac{1}{N_{trig}^{evt}}\frac{1}{\Delta R_L}\sum_{|R_{L,ij}-R_L|<\Delta R_L/2}\frac{p_{T,i}p_{T,j}}{p_{T,tot}^2}.
%&\times \Theta(R_{L,ij}-(R_L-\frac{\Delta R_L}{2})) \\
%&\times \Theta((R_L+\frac{\Delta R_L}{2})-R_{L,ij})
\end{split}
\end{equation}
Here $\Delta R_L$ represents the bin size for the angular distance between the selected hadron pairs. This analysis receives contribution from all selected particles within the Toward reigon cone, therefore suffering from main hard parton radiation signal (S) in the event and the background (B) components of the underlying event simultaneously~\cite{Field:2001aok,CDF:2004jod}. The particle pairs analyzed in the Toward cone region EEC can be regarded as the convolution of both signal and background particle correlations $(S+B)\otimes(S+B)$. To subtract the background contribution in this EEC function, dedicated correlation pairs can be made with the following strategic formula
\begin{eqnarray}\label{eqn:signal_EEC}
S\otimes S = (S+B)\otimes (S+B) \nonumber \\
+ B\otimes B  -2 (S+B)\otimes B.
\end{eqnarray}
In this equation, S means signal, B means background. To estimate the background contribution, we introduce two Transverse cones with the same $R^{cone}$ as Toward region cone perpendicular to the leading particle in azimuthal space, as shown in Fig.~\ref{fig:different_cone}. 
Assuming that the background particles from the underlying event is homogeneous and dominant in the Transverse cone region, the background contribution in the Toward cone EEC can be subtracted by designing particle correlations employing charged hadrons satisfying $p_T>1$ GeV$/c$ in the Transverse cones. The first two terms on the right hand side of Eq.~\ref{eqn:signal_EEC} can be obtained with the particle pairs within the Toward cone and two Transverse cones, respectively. The last term is then calculated using the selected particles in each Transverse cone paired with the Toward cone particles rotated by $\pm \pi/2$ to coincide the particles from different cone areas. The formula in Eq.~\ref{eqn:signal_EEC} is then translated to the Toward and Transverse cone particle combinations as:
\begin{equation}\label{eqn:sig_EEC_region}
\begin{aligned}[b]
N_{\mathrm{EEC}}^{sub} &= N_{\mathrm{EEC}}^{Tow} + N_{\mathrm{EEC}}^{Trans} - 2 N_{\mathrm{EEC}}^{Tow\otimes Trans}\\
&= N_{\mathrm{EEC}}^{Tow} - N_{\mathrm{EEC}}^{Bkg}.
\end{aligned}
\end{equation}
Here, $N_{\mathrm{EEC}}^{Tow}$ is the EEC pairs in the Toward cone, $N_{\mathrm{EEC}}^{Trans}$ is the EEC pairs in the Transverse cone, $N_{\mathrm{EEC}}^{Tow\otimes Trans}$ is the EEC pairs between the Toward cone and the Transverse cone. And we define the last two terms as the EEC Background $ N_{\mathrm{EEC}}^{Bkg}=2N_{\mathrm{EEC}}^{Tow\otimes Trans}-N_{\mathrm{EEC}}^{Trans}$. The particles from two Transverse cones are averaged to account for the impact of cone numbers.

\begin{figure}[hbt!]
	\includegraphics[width=1.0\linewidth]{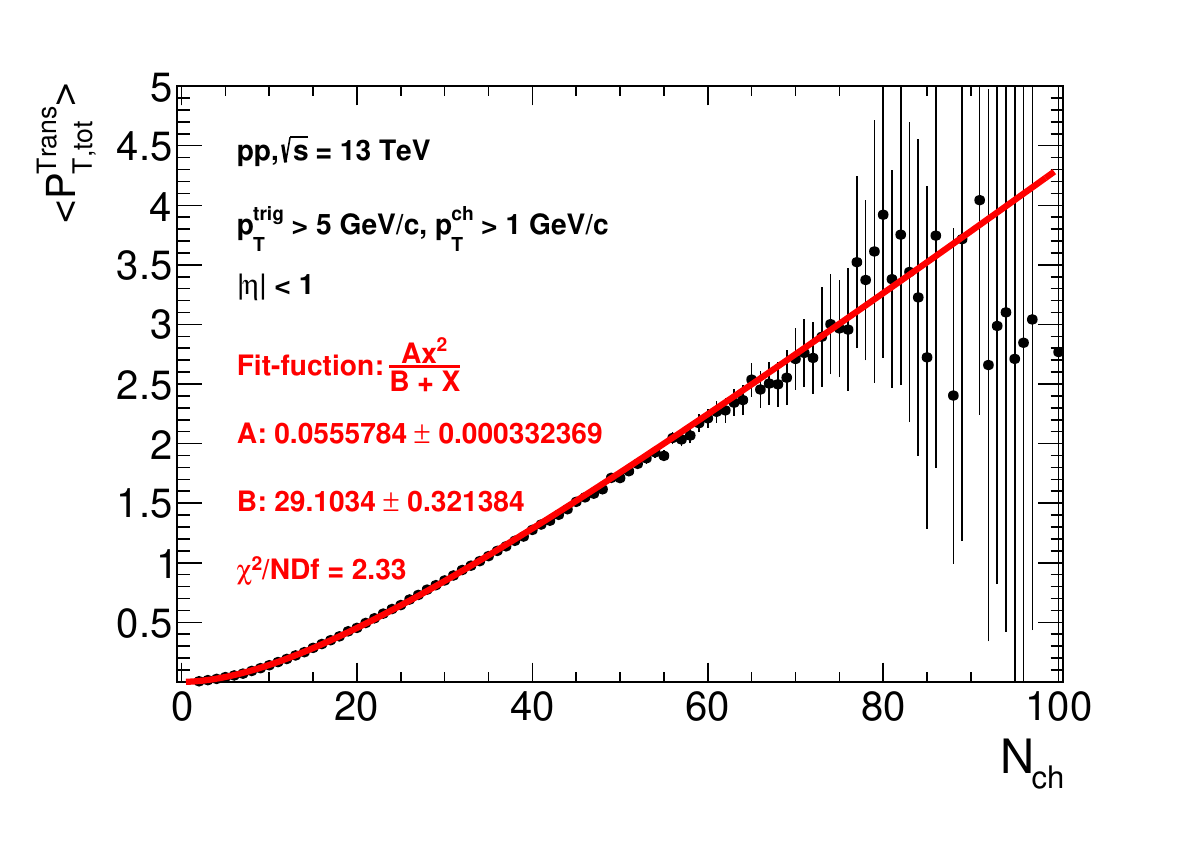}
	\caption{\label{fig:NchvsTransAvg}The average of sum $p_T$ in the Transverse cone as the function of $N_{ch}$. 
	}
\end{figure}
%the denominator correction
Considering that $p_{T,tot}$ also receives the signal contribution and background contribution in the Toward cone, we employ the average scalar sum of $p_T$ for particles involved in the EEC calculation in each Transverse cone $\langle p_{T,tot}^{Trans}\rangle$ to correct the $p_{T,tot}^{Tow}$. Since the mean value of the sum $p_T$ in the underlying event is related to the multiplicity, we provide a parameterized fit to the $\langle p_{T,tot}^{Trans}\rangle$ dependent on $N_{ch}$ estimated charge particle number with $p_T>0.4$ GeV/$c$ within $|\eta|<0.5$ in Fig.~\ref{fig:NchvsTransAvg}. The corrected $p_{T,tot}^{corr}$ is obtained by subtracting the multiplicity dependent background $p_T$ contribution averaged over multiple events
\begin{equation}\label{eqn:pT_EEC_sub}
\begin{aligned}[b]
p_{T,tot}^{corr} = p_{T,tot}^{Tow} - \langle p_{T,tot}^{Trans}\rangle.
\end{aligned}
\end{equation}
As $p_{T,tot}^{Tow}$ is significantly larger than $\langle p_{T,tot}^{Trans}\rangle$ within the accessible multiplicity region in pp collisions, this corrected $p_{T,tot}^{corr}$ is expected to be positive definite.

\section{Results and Discussion}\label{sec:results}
The simulation in this work is performed for the pp collisions at $\sqrt{s} = $ 13 TeV using PYTHIA8 Monte Carlo generator~\cite{Bierlich:2022pfr}.
The cone size used to make the correlated pairs has been fixed to $R^{cone}=0.5$. The leading particles are required to to be a charged particle with $ p_{T}>$ 5 GeV$/c$ and $|\eta|<1$ if specified otherwise in the rest of this work. Charged particles within the Toward and Transverse cones satisfying $ p_{T}>$ 1 GeV/$c$ and $|\eta|<1$ are used to make the EEC pairs. 
The resulting EEC distribution is normalized by the number of qualifying trigger events containing a valid leading particle. This normalization preserves the intrinsic shape and scaling behavior of the correlation function while enabling a direct comparison of the absolute yield between different measurements.

\begin{figure}[hbt]
	\includegraphics[width=1.0\linewidth]{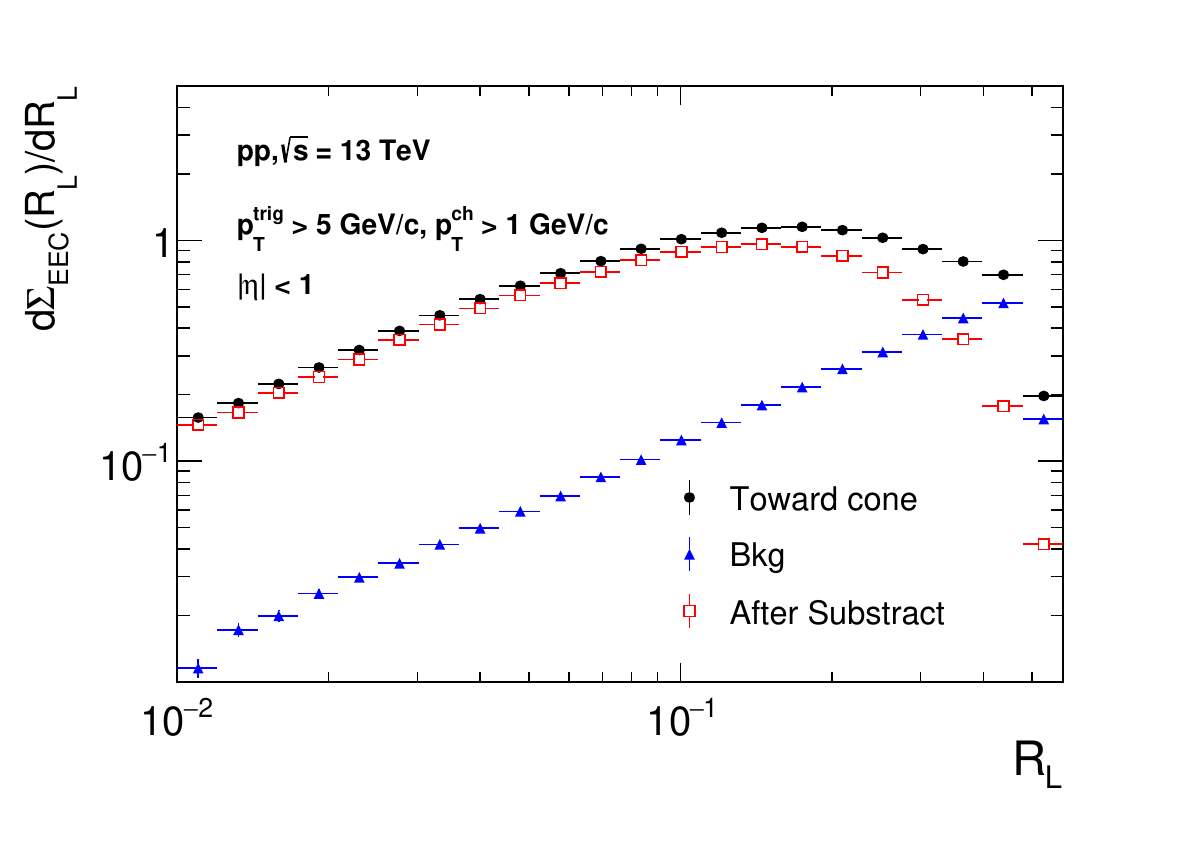}
	\caption{\label{fig:bkg_Subtract}The Toward cone EEC distribution before subtracting the background (black solid circles), the background EEC distribution defined in Eq.~\ref{eqn:sig_EEC_region} (blue solid triangle), the EEC distribution after the background subtraction (red open squares). }
\end{figure}

To understand the background subtraction procedure and its effect on the EEC distribution, we present the raw EEC distribution using pairs from Toward cone, background distributions estimated using Transverse cone together with the combination between rotated Toward cone particles and Transverse cone particles, and the final background subtracted EEC distribution in Fig.~\ref{fig:bkg_Subtract}. The raw EEC, measured in the Toward cone, indicated by the solid black dots, represents the $(S+B)\otimes(S+B)$ correlations from both the jet signal and the underlying event background. The background contribution, obtained with the Transverse cone and rotated Toward cone following the prescription in $N_{\mathrm{EEC}}^{Bkg}$ of Eq.~\ref{eqn:sig_EEC_region}, is indicated by the solid blue triangles. It is shown in Fig.~\ref{fig:bkg_Subtract} that the contribution from the background increases approximately linearly with angular distance $R_L$, then falls rapidly near the cone edge ($R_L \sim R^{\text{cone}}$), consistent with a homogeneous underlying event expectation dominated by MPI process. While negligible at small $R_L$, the background constitutes a substantial fraction of the raw yield of Toward cone EEC for $R_L \gtrsim R^{\text{cone}}$. The final EEC signal after background subtraction is shown by the red open squares. The purified EEC distribution exhibits characteristic geometric scaling at small $R_L$, followed by a radiation plateau at intermediate angular separations, and a rapid decline as $R_L$ approaches and exceeds the cone boundary.

\begin{figure}[bht]
\includegraphics[width=1.0\linewidth]{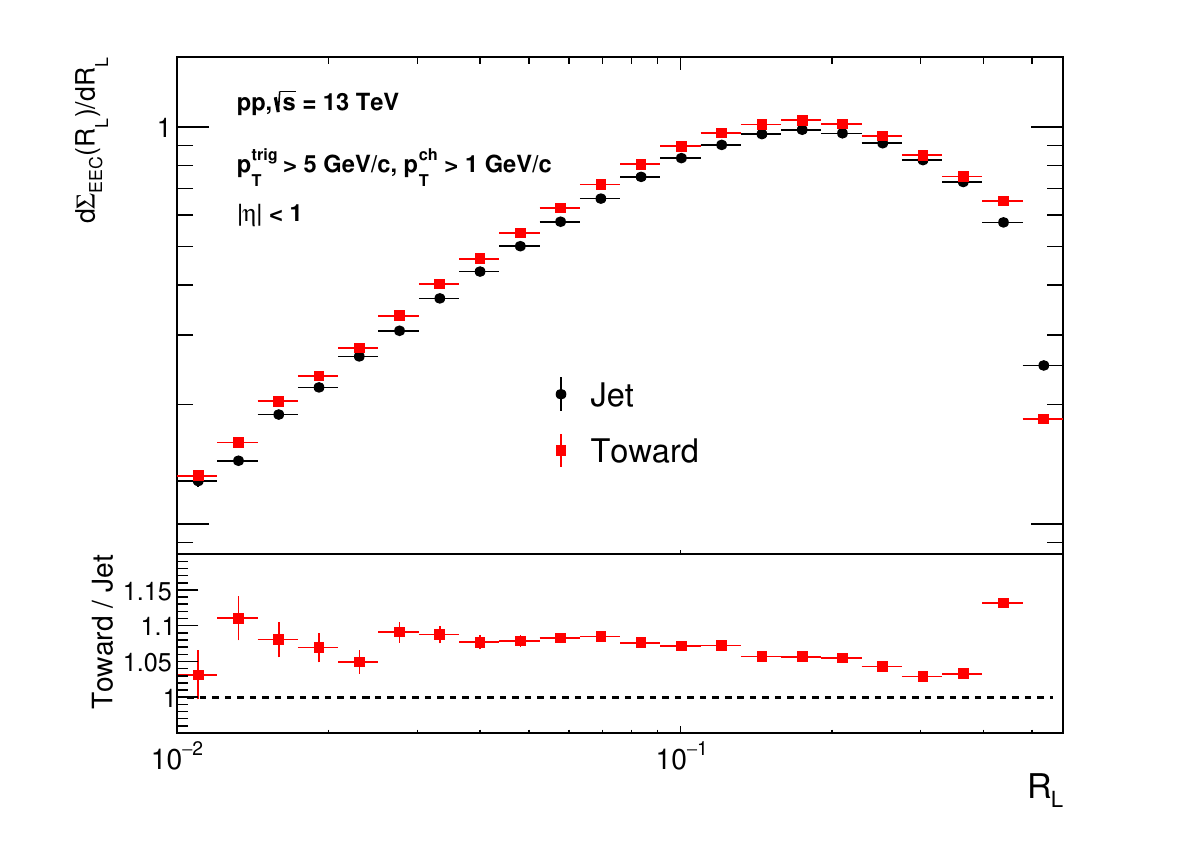}
\caption{\label{fig:different_jetconstrution}hadron-based method EEC distribution compare with jet-reconstructed-based EEC distribution.}
\end{figure}

%Briefly describe the standard jet-reconstruction method used for comparison
To validate the proposed hadron-based approach, we compare it against results obtained from full jet reconstruction, as illustrated in Fig.~\ref{fig:different_jetconstrution}.
For the same event sample, requiring a leading charged particle with $p_{T} > 5$~GeV/$c$, jets are reconstructed using the anti-$k_{t}$ algorithm with a resolution parameter of $R = 0.5$. The input to the clustering algorithm consists of charged hadrons with $p_{T} > 0.15$~GeV/$c$ within the pseudorapidity range $|\eta| < 1$. The jet axis and kinematics are determined using the $E$-scheme recombination method~\cite{Cacciari:2011ma}, summing over all constituent particles within the jet.
After substituting the Toward cone $p_{T,tot}$ and its associated charged particles in Eq.~\ref{eqn:theory_EEC} with the reconstructed leading jet $p_T$ and the jet constituent particles with $p_T>1$ GeV$/c$, we obtain the leading jet EEC with full jet reconstruction method shown by the black points in Fig.~\ref{fig:different_jetconstrution}. The EEC distribution obtained with the jet reconstruction method is found to be very similar to the Toward cone hadron method EEC shown by the red points, demonstrating that the hadron proxy in the Toward cone effectively captures the core energy flow of the jet. Discrepancies observed in the comparison especially at large $R_L$ are expected to come from the differences in the geometric treatment of the jet boundary and the recovered jet energy between the two methods. While the Toward cone method utilizes a uniform radius centered on the leading hadron, the jet algorithm produces a dynamic boundary that depends on the local distribution of radiation. The overall consistency between the two methods confirms the robustness of the hadron-based approach for probing jet substructure.

The consistency between the hadron-based method and conventional jet reconstruction can be understood from the intrinsic energy flow topology based on the leading hadron approach. By centering the Toward cone on the highest-$p_T$ particle, the hadron-based method effectively defines a reference axis analogous to the Winner-Take-All (WTA) recombination scheme~\cite{Larkoski:2014uqa,Cacciari:2014gra,Cal:2019gxa}, where the jet axis follows the most energetic constituent rather than the momentum weighted 
$E$-scheme axis. This correspondence is particularly important for EEC measurements, since the WTA axis is insensitive to soft, wide-angle radiation. Using the leading hadron as a proxy for the jet axis therefore provides a track-based definition of the energy-flow direction. The agreement observed in Fig.~\ref{fig:different_jetconstrution} demonstrates that the leading-hadron axis provides an experimentally clean and theoretically well-motivated surrogate for the jet direction, especially in high-multiplicity environments where conventional jet axes are more susceptible to underlying-event fluctuations.

\begin{figure}[hbt]
\includegraphics[width=1.0\linewidth]{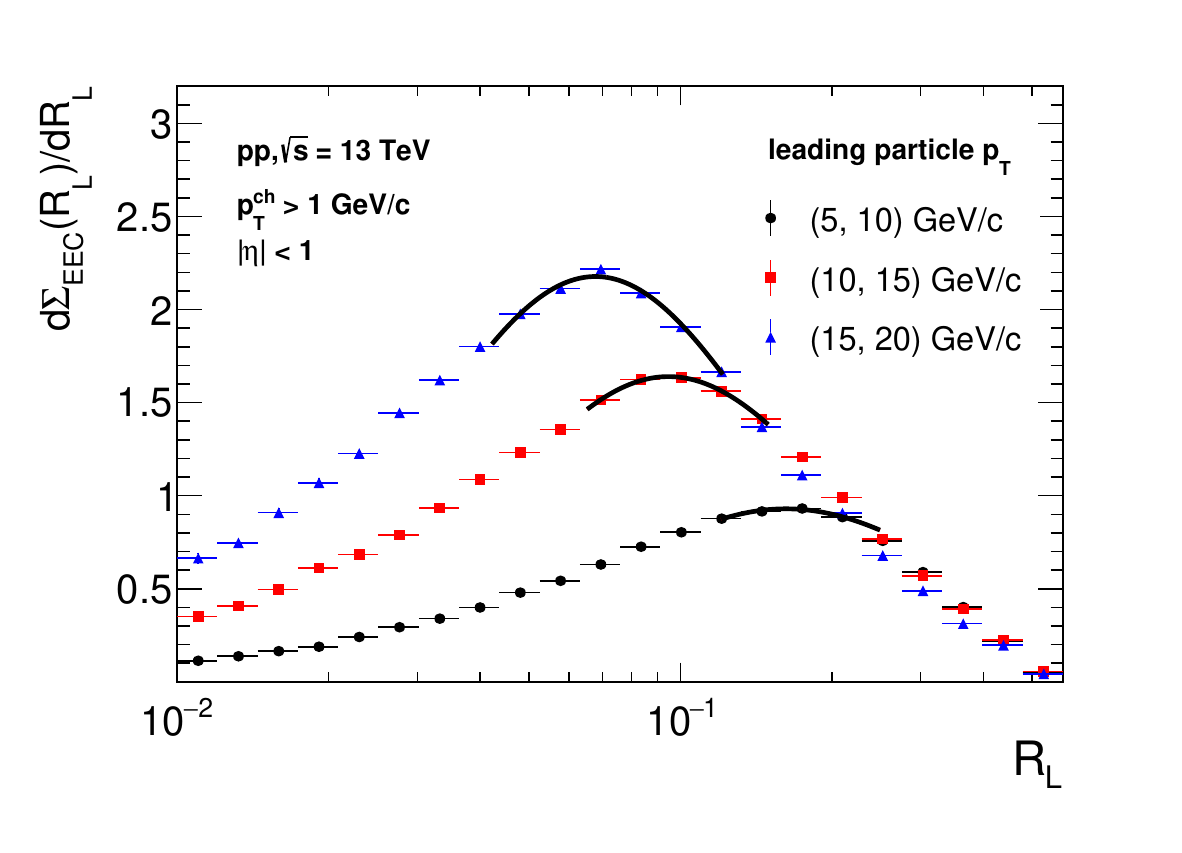}
\caption{\label{fig:trigger_pt_dependence}The EEC distributions after background subtraction applied in different trigger $p_{T}$ range.}
%fitting

\end{figure}

% discuss the scaling feature of the hadron EEC
The background-subtracted EEC distributions for leading charged hadron transverse momentum intervals of $5 - 10$ (black circles), $10 - 15$ (red squares), $15 - 20$ (blue triangles) GeV$/c$ are presented in Fig.~\ref{fig:trigger_pt_dependence}. By employing the leading particle $p_{T}$ as a proxy for jet energy scale, we investigate how the angular energy flow encoded in the EEC evolves with the underlying jet energy. The distributions exhibit a definitive transition from a small $R_{L}$ regime, where non-perturbative hadronization effects dominate, to a large $R_{L}$ regime governed by perturbative splittings. A characteristic shift of the EEC peak toward smaller angular separations has been observed as the leading particle $p_{T}$ increases. This observation is consistent with the scaling of the confinement transition that jets with higher energy undergo longer perturbative evolution, allowing the parton shower to resolve smaller angular scales before the onset of hadronization~\cite{ALICE:2024dfl}. Consequently, the peak of the EEC distribution, which marks the transition between partonic and hadronic dynamics, moves to smaller angles as the available phase space for perturbative radiation expands.

The transition between the small $R_L$ region and the large $R_L$ region provides a unique opportunity to probe the dynamics of parton-to-hadron binding, shedding light on the interplay between perturbative and non-perturbative mechanisms in QCD. To quantify this transition, we characterize the peak of the hadron-based EEC by fitting the distributions in Fig.~\ref{fig:trigger_pt_dependence} with a log-normal function across different trigger $p_T$ intervals, following the procedure in Ref.~\cite{ALICE:2024dfl}. Our analysis reveals that the peak position, $R_L^{peak}$, is inversely proportional to the corrected average transverse momentum, $< p_{T,tot}^{corr} >$, while the peak height scales linearly with $< p_{T,tot}^{corr} >$, using the values summarized in Tab.~\ref{tab:ptavg}. We observe that the product $< p_{T,tot}^{corr} > R_L^{peak}$ and the peak height normalized by $< p_{T,tot}^{corr} >$ remain approximately invariant across $p_T$ intervals, with the exception of a slight deviation in the lowest $p_T$ bin. This discrepancy likely reflects the broader, more diverse angular energy distributions of low-energy jets, which may not be fully captured by the fixed cone size employed in this analysis. However, if the $< p_{T,tot}^{corr} >$ for this lowest bin is scaled by a factor of 1.15, the resulting peak height/$< p_{T,tot}^{corr} >$ ratio and the product $< p_{T,tot}^{corr} > R_L^{peak}$ align closely with the values seen across all other bins.

\begin{table}[hbt!]
 \renewcommand{\arraystretch}{1.2}
\centering
\begin{tabular}{c|c|c|c} 
\hline
$p_{T}^{trigger}$(GeV/c) & \textbf{(5,10)}  & \textbf{(10,15)} & \textbf{(15,20)}   \\ \hline
$R_{L}^{peak} $ & 0.162 & 0.095 & 0.067 \\ \hline
EEC height  &  0.928 & 1.640 & 2.188 \\ \hline
$<p_{T,tot}^{corr}>$ & 9.361 & 18.81 & 25.65 \\ \hline
$<p_{T,tot}^{corr}>*R_{L}^{peak} $ & 1.516  & 1.787 & 1.719 \\ \hline
EEC height/$<p_{T,tot}^{corr}>$ & 0.099 & 0.087 & 0.085 \\ \hline
\end{tabular}
\caption{\label{tab:ptavg} The mean $p_{T,tot}^{corr}$, the $R_{L}^{peak} $,the EEC height in different leading particle range.}
\end{table}

\begin{figure}[hbt]
\includegraphics[width=1.0\linewidth]{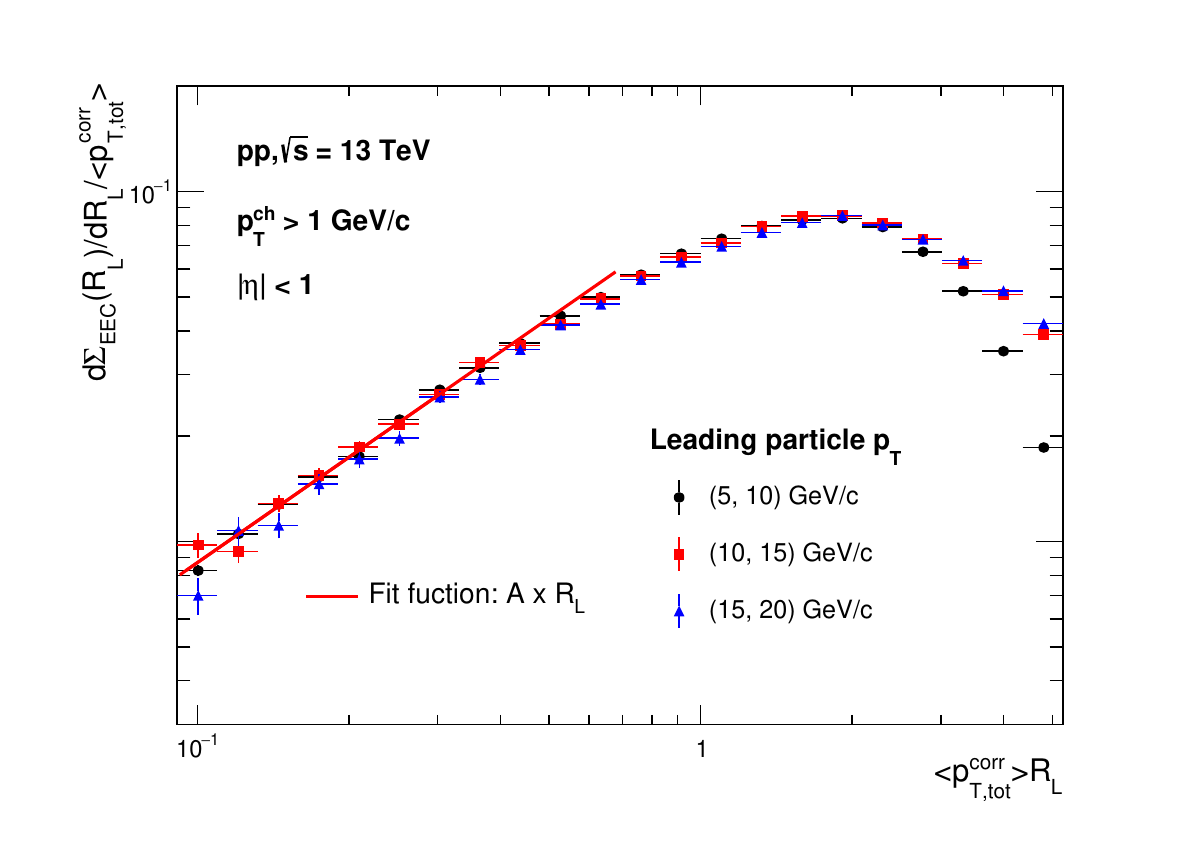}
\caption{\label{fig:scaling}The scaling behavior EEC in different leading particle $p_{T}$ range. The corrected EEC function as a function of $R_L$ has been rescaled by the mean corrected sum transverse momentum $<p_{T,tot}^{corr}>$ in each leading particle $p_{T}$ bin. The $<p_{T,tot}^{corr}>$ in the 5–10 GeV/c bin is multiplied by 1.15.}
\end{figure}
To directly visualize the scaling feature of the EEC observable, the scaled EEC distributions as a function of $<p_{T,tot}^{corr}> R_{L}$ is presented in Fig.~\ref{fig:scaling} in $p_T$ of $5 - 10$ (black circles), $10 - 15$ (red squares), $15 - 20$ GeV$/c$ (blue triangles) range. The scaled EEC function for different $p_T$ bins collapse into a common curve, after the 1.15 scaling factor for $< p_{T,tot}^{corr} >$ of the lowest $p_T$ bin is considered. A proximal scaling behavior can be found for this scaled EEC function across different leading hadron $p_T$ bins. Furthermore, it exhibits linear behavior in the small $R_{L}$ region, which may be related to the characteristics of the non-perturbative free hadron scaling depending only on the area of the infintesimal $R_L$~\cite{Komiske:2022enw,CMS:2024mlf}. 
We can observe from the Fig.~\ref{fig:scaling} that EEC with the leading hadron $p_T$ within $5-10$ GeV/c shows slight deviation at large $R_L$. This breakdown may be attributed to the hadronization dynamics in this softer momentum regime, where the collimation between the trigger hadron and the initiating parton is reduced, and the proportionality of summed total transverse momentum is more significantly influenced by underlying-event background.

%Parton Matching
%Introduce the method used to match leading particles to their mother partons (quarks vs. gluons) in the simulation.
The EEC provides a powerful framework for probing the flavor-dependent dynamics of parton showers and hadronization, extending beyond its role in characterizing general angular structure~\cite{Apolinario:2025vtx,Chen:2024quk}. Its sensitivity originates from fundamental differences in color charge and radiation patterns between quark and gluon jets, which manifest as distinct shapes and scaling behaviors in the EEC. This behavior extends naturally to the heavy-flavor sector, where the finite masses of charm and beauty quarks modify collinear evolution by suppressing small-angle radiation through the so called dead cone effect, leaving clear imprints on the EEC distribution. Consequently, once the correlation signal is cleanly extracted through the hadron-based method, this measurement enables quantitative studies of quark–gluon discrimination, dead-cone signatures, and the combined influence of parton mass and color structure on in-medium energy flow.

To investigate these flavor-dependent effects, the partonic origin of each leading hadron is determined through a geometric matching procedure within the PYTHIA8 event generator. The leading hadron is associated with its initiating parton by identifying the highest-$p_T$ parton from the primary hard-scattering process that resides within the Toward cone of radius $R_{\text{cone}}$. Events are subsequently categorized based on the flavor of the matched initiator, allowing for a systematic comparison of the EEC patterns arising from light-quark and gluon populations. This classification facilitates a quantitative assessment of how the fundamental properties of the initiating parton propagate through the fragmentation chain to the final-state energy flow, validating the hadron-based approach as a sensitive probe for testing QCD flavor dynamics in both vacuum and dense nuclear medium.

%Quark vs. Gluon EECs
%Show the distinct shapes of EECs initiated by light quarks versus gluons. Discuss how gluon-initiated EECs peak at larger RL and have a larger magnitude, consistent with theory (larger color factor, stronger radiation).
%We can explore the EEC distribution which has different mother parton by employing the mother parton matching method mentioned above. Taking the case where the leading particle is a charged particle with $ P_{T} >$5 GeV as an example, we investigate the differences in shape and scale of the different mother parton EEC.

\begin{figure}[hbt]
\includegraphics[width=1.0\linewidth]{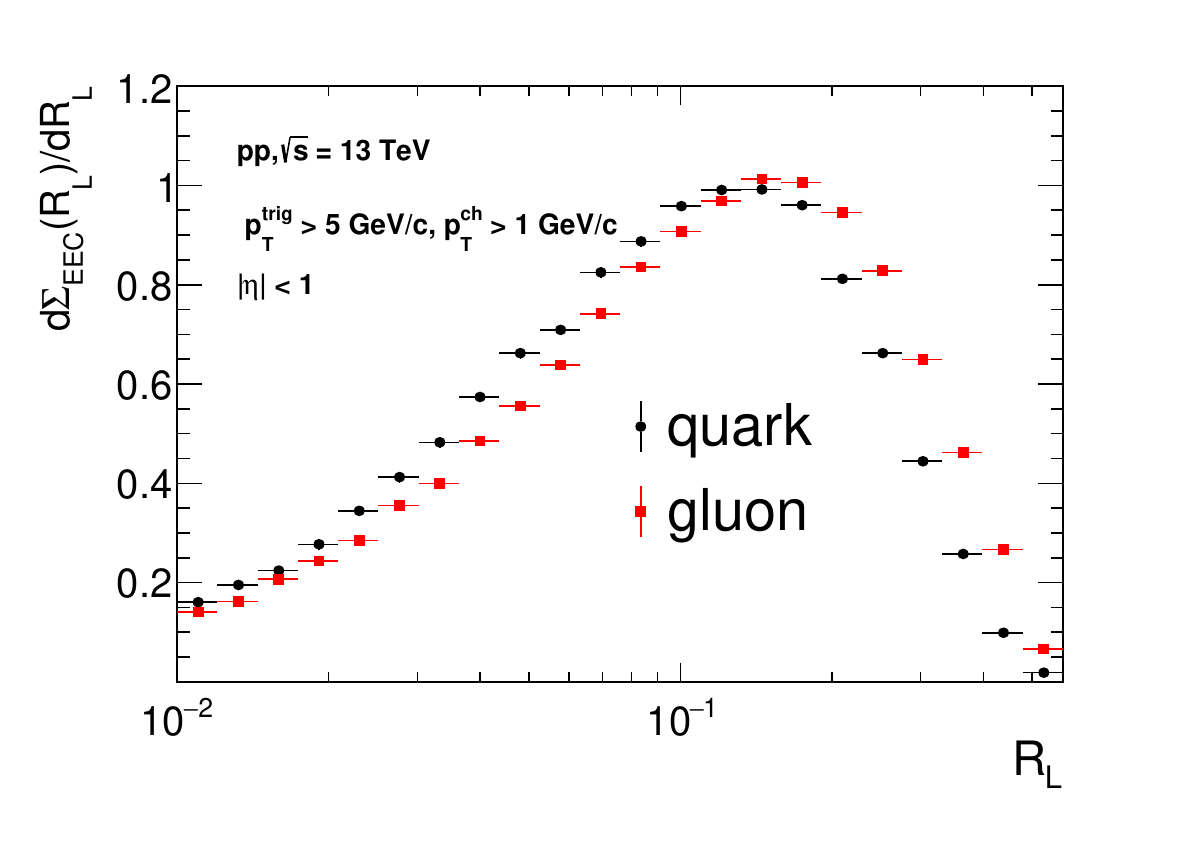}
\caption{\label{fig:quark_gluon}The EEC distributions of different mother parton after background subtraction.}
\end{figure}
Figure~\ref{fig:quark_gluon} presents a comparison of the EEC distributions for quark-initiated (black circles) and gluon-initiated (red squares) events.
A distinct dependence on the flavor of the initiating parton is observed in both the magnitude and the shape of the distribution. The gluon-initiated EEC exhibits a significantly higher overall normalization and a peak shifted toward larger angular separations ($R_{L}$) relative to the quark-initiated counterpart. This behavior is consistent with the more intense and spatially extended radiation pattern characteristic of gluon jets, as well as an earlier onset of hadronization. These differences originate from the larger Casimir color factor of gluons compared to quarks, which governs the strength and angular broadening of the parton shower in gluon-initiated jets~\cite{Forte:2025twm}.

Heavy-flavor (HF) jets provide a unique laboratory for studying mass-dependent effects in QCD. Their evolution is governed by the dead-cone effect, whereby gluon radiation is suppressed at small angles due to the finite quark mass~\cite{Dokshitzer:1991fd,ALICE:2021aqk}. Because the EEC encodes the full evolution from initial parton production through hadronization, it is particularly sensitive to such angular modifications. A comparison between HF- and light-flavor–initiated jets therefore isolates this suppression and offers a direct experimental handle on heavy-quark mass effects~\cite{Xing:2024yrb,Craft:2022kdo,ALICE:2025igw}.
\begin{figure}[h]
\includegraphics[width=1.0\linewidth]{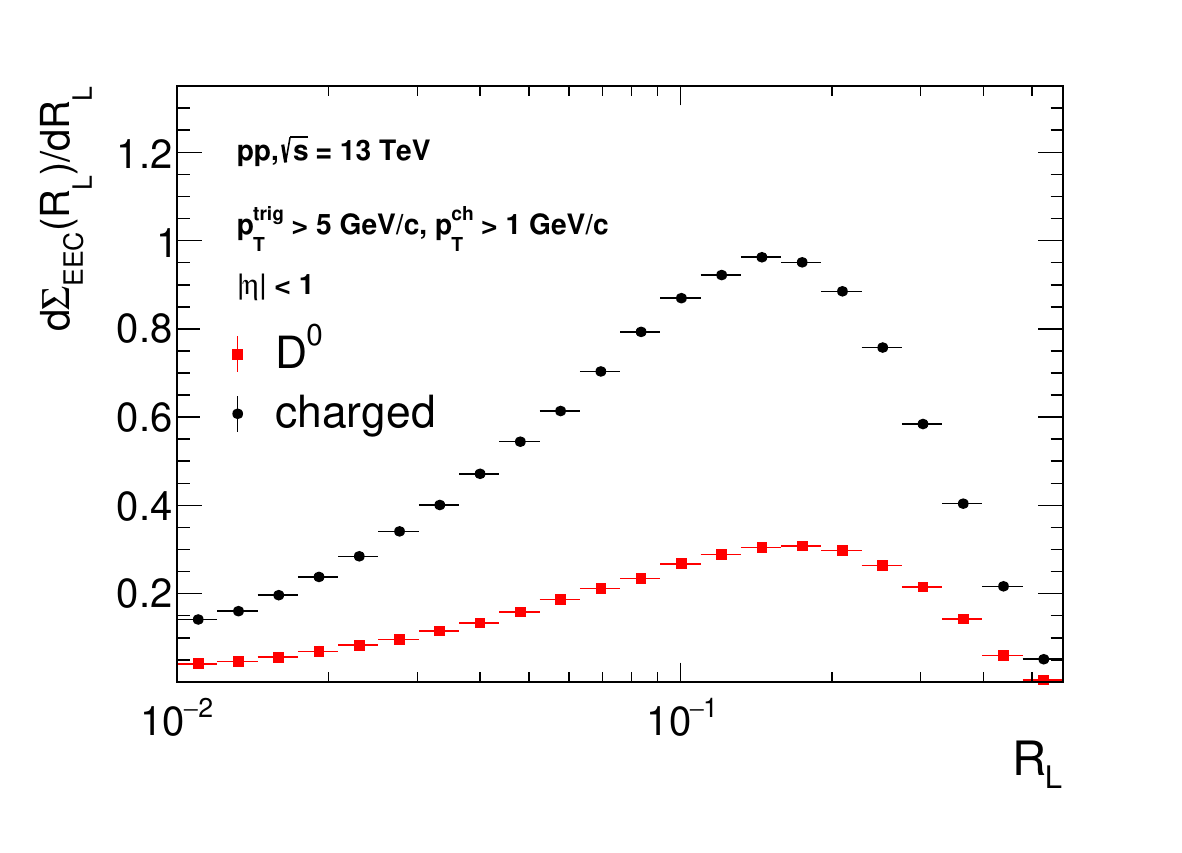}
\caption{\label{fig:flavor_dependence}The EEC distributions after background subtraction with $\rm D^{0}$ hadron (red points) and charged hadron (black points) as leading particle.}
\end{figure}
To expose this behavior, Fig.~\ref{fig:flavor_dependence} compares the EEC distributions triggered by leading $D^{0}$ mesons with those triggered by inclusive charged particles. In this framework, the leading $D^{0}$ meson serves as a proxy for charm-quark fragmentation, whereas the inclusive sample is predominantly composed of light-quark and gluon-initiated jets. We emphasize that feed-down contributions from $B$-meson decays and secondary $D$-hadron production are excluded to isolate the dynamics of prompt charm production. Relative to the inclusive charged-particle baseline, the $D^{0}$-triggered EEC exhibits a markedly reduced magnitude and a peak shifted toward larger angular separations. This observation is a clear manifestation of the dead-cone effect, in which the suppression of small-angle gluon radiation by the charm-quark mass results in a broader and less intense EEC pattern. The distinct contrast between these distributions thus provides a direct measurement of the influence of parton mass on the angular architecture of the parton shower.
Overall, the jet-free leading-hadron EEC method faithfully captures the scale and flavor dependence features of parton evolution and transition to hadronic final states, validating the hadron based EEC as a sensitive probe of the underlying QCD dynamics with direct applicability to nuclear collisions.

\section{summary}\label{sec:summary}
%Summary of Work
In this work, we have introduced a jet-free framework for measuring EEC in pp collisions using an event-topology–based method. By employing the leading charged hadron as a reference axis and utilizing Toward and Transverse regions for background subtraction, this approach avoids explicit jet reconstruction while retaining sensitivity to the essential features of perturbative parton shower evolution and non-perturbative hadronization process. Comparisons with conventional jet-based EEC measurements demonstrate strong consistency, validating the leading-hadron axis as a robust and experimentally clean proxy for the jet direction.
%Key Findings
The proposed method enables access to low-$p_T$ EEC measurements that are typically challenging for standard jet-based analyses. We observe a clear evolution of the EEC with the leading-particle transverse momentum, including a systematic shift of the EEC peak toward smaller angular separations as the hard scale increases. The approximate scaling behavior of the peak position with the inverse leading-particle momentum indicates that the leading hadron provides an effective measure of the jet energy scale and that the EEC peak traces the transition between perturbative parton evolution and non-perturbative hadronization. The emergence of linear behavior at small angular separations further highlights the sensitivity of this observable to hadronization dynamics.

Beyond scale dependence, the jet-free EEC framework exhibits strong sensitivity to parton flavor. Distinct differences are observed between quark- and gluon-initiated events, reflecting their different color charges and radiation patterns. Extending the analysis to heavy flavor, we show that EEC triggered by leading $D$ meson mesons display a suppressed magnitude and a peak shifted toward larger angular separations relative to inclusive charged-particle triggers. These features provide a clear manifestation of the dead-cone effect, demonstrating the ability of the EEC to directly probe mass-dependent QCD dynamics within a hadron-based approach. 
%Future Outlook
%●suggest potential applications, such as applying the method to experimental data from pp or heavy-ion collisions to provide new constraints on hadronization models.
Looking forward, this method is well suited for application to experimental data in pp collisions, where it can provide new constraints on hadronization models and parton-shower dynamics without reliance on full jet reconstruction. Its robustness and simplicity make it particularly attractive for heavy-ion and proton–nucleus collisions, where large background fluctuations complicate traditional jet analyses. In such environments, the jet-free EEC offers a promising avenue to study medium-induced modifications of energy flow, including soft radiation, jet broadening, and medium response effects, especially in the low- to intermediate-$p_T$ regime. Future extensions to more differential measurements, such as multiplicity-dependent EEC studies, will further establish the jet-free EEC as a versatile and powerful probe of QCD dynamics across different collision systems.

\section*{}
\begin{acknowledgments}
This work was supported by the National Key Research and Development Program of China (Grant No. 2024YFA1610803), the National Natural Science Foundation of China (Grant Nos. 12275103 and W2543004), the Fundamental Research Funds for the Central Universities, China University of Geosciences(Wuhan) with No. G1323523064, and the Innovation Fund of Key Laboratory of Quark and Lepton Physics QLPL2025P01.
\end{acknowledgments}

%\appendix

%\section{Appendixes}
%%Backup to explain jet/leading pt contribution source
%\begin{figure}[ht]
%\includegraphics[width=1.0\linewidth]{Match_portion_chp.pdf}
%\caption{\label{Match_portion}Share of leading particle matched to different mother parton as a function of the leading particle $P_{T}$ with $ P_{T} > $ 0.15 GeV/c. The black circles are all leading particles that have not matched the mother parton, the red circles are the leading particles matched to light quark, blue circles are the leading particles matched to gluon.}
%\end{figure}
% From FIG.~\ref{Match_portion}, it can be observed that the probability of the leading particle come from different mother parton, whether is a charged particle or a charged pion (here we just exhibit the charged particle), The matching probability exceeds 80\% . the share of quark increases with the increase of leading particle $P_{T}$, and eventually also tends to be stable, while the share of gluon firstly reaches a peak around 2 GeV/c and finally decreases, eventually also tends to stable. The gluons outweigh the quarks. When the leading particle is a proton or a heavy particle,it will present a little different situation, but we don't show here individually.
%

\nocite{*}

\bibliography{apssamp}% Produces the bibliography via BibTeX.

\begin{thebibliography}{47}%
\makeatletter
\providecommand \@ifxundefined [1]{%
 \@ifx{#1\undefined}
}%
\providecommand \@ifnum [1]{%
 \ifnum #1\expandafter \@firstoftwo
 \else \expandafter \@secondoftwo
 \fi
}%
\providecommand \@ifx [1]{%
 \ifx #1\expandafter \@firstoftwo
 \else \expandafter \@secondoftwo
 \fi
}%
\providecommand \natexlab [1]{#1}%
\providecommand \enquote  [1]{``#1''}%
\providecommand \bibnamefont  [1]{#1}%
\providecommand \bibfnamefont [1]{#1}%
\providecommand \citenamefont [1]{#1}%
\providecommand \href@noop [0]{\@secondoftwo}%
\providecommand \href [0]{\begingroup \@sanitize@url \@href}%
\providecommand \@href[1]{\@@startlink{#1}\@@href}%
\providecommand \@@href[1]{\endgroup#1\@@endlink}%
\providecommand \@sanitize@url [0]{\catcode `\\12\catcode `\$12\catcode
  `\&12\catcode `\#12\catcode `\^12\catcode `\_12\catcode `\%12\relax}%
\providecommand \@@startlink[1]{}%
\providecommand \@@endlink[0]{}%
\providecommand \url  [0]{\begingroup\@sanitize@url \@url }%
\providecommand \@url [1]{\endgroup\@href {#1}{\urlprefix }}%
\providecommand \urlprefix  [0]{URL }%
\providecommand \Eprint [0]{\href }%
\providecommand \doibase [0]{https://doi.org/}%
\providecommand \selectlanguage [0]{\@gobble}%
\providecommand \bibinfo  [0]{\@secondoftwo}%
\providecommand \bibfield  [0]{\@secondoftwo}%
\providecommand \translation [1]{[#1]}%
\providecommand \BibitemOpen [0]{}%
\providecommand \bibitemStop [0]{}%
\providecommand \bibitemNoStop [0]{.\EOS\space}%
\providecommand \EOS [0]{\spacefactor3000\relax}%
\providecommand \BibitemShut  [1]{\csname bibitem#1\endcsname}%
\let\auto@bib@innerbib\@empty
%</preamble>
\bibitem [{\citenamefont {Hofman}\ and\ \citenamefont
  {Maldacena}(2008)}]{Hofman:2008ar}%
  \BibitemOpen
  \bibfield  {author} {\bibinfo {author} {\bibfnamefont {D.~M.}\ \bibnamefont
  {Hofman}}\ and\ \bibinfo {author} {\bibfnamefont {J.}~\bibnamefont
  {Maldacena}},\ }\bibfield  {title} {\bibinfo {title} {{Conformal collider
  physics: Energy and charge correlations}},\ }\href
  {https://doi.org/10.1088/1126-6708/2008/05/012} {\bibfield  {journal}
  {\bibinfo  {journal} {JHEP}\ }\textbf {\bibinfo {volume} {05}},\ \bibinfo
  {pages} {012}},\ \Eprint {https://arxiv.org/abs/0803.1467} {arXiv:0803.1467
  [hep-th]} \BibitemShut {NoStop}%
\bibitem [{\citenamefont {Larkoski}\ \emph {et~al.}(2020)\citenamefont
  {Larkoski}, \citenamefont {Moult},\ and\ \citenamefont
  {Nachman}}]{Larkoski:2017jix}%
  \BibitemOpen
  \bibfield  {author} {\bibinfo {author} {\bibfnamefont {A.~J.}\ \bibnamefont
  {Larkoski}}, \bibinfo {author} {\bibfnamefont {I.}~\bibnamefont {Moult}},\
  and\ \bibinfo {author} {\bibfnamefont {B.}~\bibnamefont {Nachman}},\
  }\bibfield  {title} {\bibinfo {title} {{Jet Substructure at the Large Hadron
  Collider: A Review of Recent Advances in Theory and Machine Learning}},\
  }\href {https://doi.org/10.1016/j.physrep.2019.11.001} {\bibfield  {journal}
  {\bibinfo  {journal} {Phys. Rept.}\ }\textbf {\bibinfo {volume} {841}},\
  \bibinfo {pages} {1} (\bibinfo {year} {2020})},\ \Eprint
  {https://arxiv.org/abs/1709.04464} {arXiv:1709.04464 [hep-ph]} \BibitemShut
  {NoStop}%
\bibitem [{\citenamefont {Chen}\ \emph {et~al.}(2020)\citenamefont {Chen},
  \citenamefont {Moult}, \citenamefont {Zhang},\ and\ \citenamefont
  {Zhu}}]{Chen:2020vvp}%
  \BibitemOpen
  \bibfield  {author} {\bibinfo {author} {\bibfnamefont {H.}~\bibnamefont
  {Chen}}, \bibinfo {author} {\bibfnamefont {I.}~\bibnamefont {Moult}},
  \bibinfo {author} {\bibfnamefont {X.}~\bibnamefont {Zhang}},\ and\ \bibinfo
  {author} {\bibfnamefont {H.~X.}\ \bibnamefont {Zhu}},\ }\bibfield  {title}
  {\bibinfo {title} {{Rethinking jets with energy correlators: Tracks,
  resummation, and analytic continuation}},\ }\href
  {https://doi.org/10.1103/PhysRevD.102.054012} {\bibfield  {journal} {\bibinfo
   {journal} {Phys. Rev. D}\ }\textbf {\bibinfo {volume} {102}},\ \bibinfo
  {pages} {054012} (\bibinfo {year} {2020})},\ \Eprint
  {https://arxiv.org/abs/2004.11381} {arXiv:2004.11381 [hep-ph]} \BibitemShut
  {NoStop}%
\bibitem [{\citenamefont {Moult}\ and\ \citenamefont
  {Zhu}(2025)}]{Moult:2025nhu}%
  \BibitemOpen
  \bibfield  {author} {\bibinfo {author} {\bibfnamefont {I.}~\bibnamefont
  {Moult}}\ and\ \bibinfo {author} {\bibfnamefont {H.~X.}\ \bibnamefont
  {Zhu}},\ }\bibfield  {title} {\bibinfo {title} {{Energy Correlators: A
  Journey From Theory to Experiment}},\ }\href@noop {} {\  (\bibinfo {year}
  {2025})},\ \Eprint {https://arxiv.org/abs/2506.09119} {arXiv:2506.09119
  [hep-ph]} \BibitemShut {NoStop}%
\bibitem [{\citenamefont {Ali}\ and\ \citenamefont
  {Barreiro}(1982)}]{Ali:1982ub}%
  \BibitemOpen
  \bibfield  {author} {\bibinfo {author} {\bibfnamefont {A.}~\bibnamefont
  {Ali}}\ and\ \bibinfo {author} {\bibfnamefont {F.}~\bibnamefont {Barreiro}},\
  }\bibfield  {title} {\bibinfo {title} {{An O ($\alpha^- s^2$) Calculation of
  Energy-energy Correlation in $e^+ e^-$ Annihilation and Comparison With
  Experimental Data}},\ }\href {https://doi.org/10.1016/0370-2693(82)90621-9}
  {\bibfield  {journal} {\bibinfo  {journal} {Phys. Lett. B}\ }\textbf
  {\bibinfo {volume} {118}},\ \bibinfo {pages} {155} (\bibinfo {year}
  {1982})}\BibitemShut {NoStop}%
\bibitem [{\citenamefont {Gao}\ \emph {et~al.}(2019)\citenamefont {Gao},
  \citenamefont {Li}, \citenamefont {Moult},\ and\ \citenamefont
  {Zhu}}]{Gao:2019ojf}%
  \BibitemOpen
  \bibfield  {author} {\bibinfo {author} {\bibfnamefont {A.}~\bibnamefont
  {Gao}}, \bibinfo {author} {\bibfnamefont {H.~T.}\ \bibnamefont {Li}},
  \bibinfo {author} {\bibfnamefont {I.}~\bibnamefont {Moult}},\ and\ \bibinfo
  {author} {\bibfnamefont {H.~X.}\ \bibnamefont {Zhu}},\ }\bibfield  {title}
  {\bibinfo {title} {{Precision QCD Event Shapes at Hadron Colliders: The
  Transverse Energy-Energy Correlator in the Back-to-Back Limit}},\ }\href
  {https://doi.org/10.1103/PhysRevLett.123.062001} {\bibfield  {journal}
  {\bibinfo  {journal} {Phys. Rev. Lett.}\ }\textbf {\bibinfo {volume} {123}},\
  \bibinfo {pages} {062001} (\bibinfo {year} {2019})},\ \Eprint
  {https://arxiv.org/abs/1901.04497} {arXiv:1901.04497 [hep-ph]} \BibitemShut
  {NoStop}%
\bibitem [{\citenamefont {Aboona}\ \emph {et~al.}(2025)\citenamefont {Aboona}
  \emph {et~al.}}]{STAR:2025jut}%
  \BibitemOpen
  \bibfield  {author} {\bibinfo {author} {\bibfnamefont {B.~E.}\ \bibnamefont
  {Aboona}} \emph {et~al.} (\bibinfo {collaboration} {STAR}),\ }\bibfield
  {title} {\bibinfo {title} {{Measurement of Two-Point Energy Correlators
  within Jets in p+p Collisions at s=200{\,}{\,}GeV}},\ }\href
  {https://doi.org/10.1103/wv2t-dkgn} {\bibfield  {journal} {\bibinfo
  {journal} {Phys. Rev. Lett.}\ }\textbf {\bibinfo {volume} {135}},\ \bibinfo
  {pages} {111901} (\bibinfo {year} {2025})},\ \Eprint
  {https://arxiv.org/abs/2502.15925} {arXiv:2502.15925 [hep-ex]} \BibitemShut
  {NoStop}%
\bibitem [{\citenamefont {Komiske}\ \emph {et~al.}(2023)\citenamefont
  {Komiske}, \citenamefont {Moult}, \citenamefont {Thaler},\ and\ \citenamefont
  {Zhu}}]{Komiske:2022enw}%
  \BibitemOpen
  \bibfield  {author} {\bibinfo {author} {\bibfnamefont {P.~T.}\ \bibnamefont
  {Komiske}}, \bibinfo {author} {\bibfnamefont {I.}~\bibnamefont {Moult}},
  \bibinfo {author} {\bibfnamefont {J.}~\bibnamefont {Thaler}},\ and\ \bibinfo
  {author} {\bibfnamefont {H.~X.}\ \bibnamefont {Zhu}},\ }\bibfield  {title}
  {\bibinfo {title} {{Analyzing N-Point Energy Correlators inside Jets with CMS
  Open Data}},\ }\href {https://doi.org/10.1103/PhysRevLett.130.051901}
  {\bibfield  {journal} {\bibinfo  {journal} {Phys. Rev. Lett.}\ }\textbf
  {\bibinfo {volume} {130}},\ \bibinfo {pages} {051901} (\bibinfo {year}
  {2023})},\ \Eprint {https://arxiv.org/abs/2201.07800} {arXiv:2201.07800
  [hep-ph]} \BibitemShut {NoStop}%
\bibitem [{\citenamefont {Acharya}\ \emph {et~al.}(2024)\citenamefont {Acharya}
  \emph {et~al.}}]{ALICE:2024dfl}%
  \BibitemOpen
  \bibfield  {author} {\bibinfo {author} {\bibfnamefont {S.}~\bibnamefont
  {Acharya}} \emph {et~al.} (\bibinfo {collaboration} {ALICE}),\ }\bibfield
  {title} {\bibinfo {title} {{Exposing the parton-hadron transition within jets
  with energy-energy correlators in pp collisions at $\sqrt{\textit s}=5.02$
  TeV}},\ }\href@noop {} {\  (\bibinfo {year} {2024})},\ \Eprint
  {https://arxiv.org/abs/2409.12687} {arXiv:2409.12687 [hep-ex]} \BibitemShut
  {NoStop}%
\bibitem [{\citenamefont {Lee}\ \emph {et~al.}(2025)\citenamefont {Lee},
  \citenamefont {Me{\c{c}}aj},\ and\ \citenamefont {Moult}}]{Lee:2022uwt}%
  \BibitemOpen
  \bibfield  {author} {\bibinfo {author} {\bibfnamefont {K.}~\bibnamefont
  {Lee}}, \bibinfo {author} {\bibfnamefont {B.}~\bibnamefont {Me{\c{c}}aj}},\
  and\ \bibinfo {author} {\bibfnamefont {I.}~\bibnamefont {Moult}},\ }\bibfield
   {title} {\bibinfo {title} {{Conformal collider physics meets LHC data}},\
  }\href {https://doi.org/10.1103/PhysRevD.111.L011502} {\bibfield  {journal}
  {\bibinfo  {journal} {Phys. Rev. D}\ }\textbf {\bibinfo {volume} {111}},\
  \bibinfo {pages} {L011502} (\bibinfo {year} {2025})},\ \Eprint
  {https://arxiv.org/abs/2205.03414} {arXiv:2205.03414 [hep-ph]} \BibitemShut
  {NoStop}%
\bibitem [{\citenamefont {Hatta}\ \emph {et~al.}(2008)\citenamefont {Hatta},
  \citenamefont {Iancu},\ and\ \citenamefont {Mueller}}]{Hatta:2008tx}%
  \BibitemOpen
  \bibfield  {author} {\bibinfo {author} {\bibfnamefont {Y.}~\bibnamefont
  {Hatta}}, \bibinfo {author} {\bibfnamefont {E.}~\bibnamefont {Iancu}},\ and\
  \bibinfo {author} {\bibfnamefont {A.~H.}\ \bibnamefont {Mueller}},\
  }\bibfield  {title} {\bibinfo {title} {{Jet evolution in the N=4 SYM plasma
  at strong coupling}},\ }\href {https://doi.org/10.1088/1126-6708/2008/05/037}
  {\bibfield  {journal} {\bibinfo  {journal} {JHEP}\ }\textbf {\bibinfo
  {volume} {05}},\ \bibinfo {pages} {037}},\ \Eprint
  {https://arxiv.org/abs/0803.2481} {arXiv:0803.2481 [hep-th]} \BibitemShut
  {NoStop}%
\bibitem [{\citenamefont {Brown}\ and\ \citenamefont
  {Ellis}(1981)}]{Brown:1981jv}%
  \BibitemOpen
  \bibfield  {author} {\bibinfo {author} {\bibfnamefont {L.~S.}\ \bibnamefont
  {Brown}}\ and\ \bibinfo {author} {\bibfnamefont {S.~D.}\ \bibnamefont
  {Ellis}},\ }\bibfield  {title} {\bibinfo {title} {{Energy-energy Correlations
  in Electron - Positron Annihilations}},\ }\href
  {https://doi.org/10.1103/PhysRevD.24.2383} {\bibfield  {journal} {\bibinfo
  {journal} {Phys. Rev. D}\ }\textbf {\bibinfo {volume} {24}},\ \bibinfo
  {pages} {2383} (\bibinfo {year} {1981})}\BibitemShut {NoStop}%
\bibitem [{\citenamefont {Berger}\ \emph {et~al.}(1985)\citenamefont {Berger}
  \emph {et~al.}}]{PLUTO:1985yzc}%
  \BibitemOpen
  \bibfield  {author} {\bibinfo {author} {\bibfnamefont {C.}~\bibnamefont
  {Berger}} \emph {et~al.} (\bibinfo {collaboration} {PLUTO}),\ }\bibfield
  {title} {\bibinfo {title} {{A Study of Energy-energy Correlations in $e^+
  e^-$ Annihilations at $\sqrt{s}=34$.6-{GeV}}},\ }\href
  {https://doi.org/10.1007/BF01413599} {\bibfield  {journal} {\bibinfo
  {journal} {Z. Phys. C}\ }\textbf {\bibinfo {volume} {28}},\ \bibinfo {pages}
  {365} (\bibinfo {year} {1985})}\BibitemShut {NoStop}%
\bibitem [{\citenamefont {Khachatryan}\ \emph {et~al.}(2016)\citenamefont
  {Khachatryan} \emph {et~al.}}]{CMS:2015wcf}%
  \BibitemOpen
  \bibfield  {author} {\bibinfo {author} {\bibfnamefont {V.}~\bibnamefont
  {Khachatryan}} \emph {et~al.} (\bibinfo {collaboration} {CMS}),\ }\bibfield
  {title} {\bibinfo {title} {{Event generator tunes obtained from underlying
  event and multiparton scattering measurements}},\ }\href
  {https://doi.org/10.1140/epjc/s10052-016-3988-x} {\bibfield  {journal}
  {\bibinfo  {journal} {Eur. Phys. J. C}\ }\textbf {\bibinfo {volume} {76}},\
  \bibinfo {pages} {155} (\bibinfo {year} {2016})},\ \Eprint
  {https://arxiv.org/abs/1512.00815} {arXiv:1512.00815 [hep-ex]} \BibitemShut
  {NoStop}%
\bibitem [{\citenamefont {Andres}\ \emph {et~al.}(2025)\citenamefont {Andres},
  \citenamefont {Holguin}, \citenamefont {Kunnawalkam~Elayavalli},\ and\
  \citenamefont {Viinikainen}}]{Andres:2024hdd}%
  \BibitemOpen
  \bibfield  {author} {\bibinfo {author} {\bibfnamefont {C.}~\bibnamefont
  {Andres}}, \bibinfo {author} {\bibfnamefont {J.}~\bibnamefont {Holguin}},
  \bibinfo {author} {\bibfnamefont {R.}~\bibnamefont
  {Kunnawalkam~Elayavalli}},\ and\ \bibinfo {author} {\bibfnamefont
  {J.}~\bibnamefont {Viinikainen}},\ }\bibfield  {title} {\bibinfo {title}
  {{Minimizing Selection Bias in Inclusive Jets in Heavy-Ion Collisions with
  Energy Correlators}},\ }\href
  {https://doi.org/10.1103/PhysRevLett.134.082303} {\bibfield  {journal}
  {\bibinfo  {journal} {Phys. Rev. Lett.}\ }\textbf {\bibinfo {volume} {134}},\
  \bibinfo {pages} {082303} (\bibinfo {year} {2025})},\ \Eprint
  {https://arxiv.org/abs/2409.07514} {arXiv:2409.07514 [hep-ph]} \BibitemShut
  {NoStop}%
\bibitem [{\citenamefont {Yang}\ \emph {et~al.}(2024)\citenamefont {Yang},
  \citenamefont {He}, \citenamefont {Moult},\ and\ \citenamefont
  {Wang}}]{Yang:2023dwc}%
  \BibitemOpen
  \bibfield  {author} {\bibinfo {author} {\bibfnamefont {Z.}~\bibnamefont
  {Yang}}, \bibinfo {author} {\bibfnamefont {Y.}~\bibnamefont {He}}, \bibinfo
  {author} {\bibfnamefont {I.}~\bibnamefont {Moult}},\ and\ \bibinfo {author}
  {\bibfnamefont {X.-N.}\ \bibnamefont {Wang}},\ }\bibfield  {title} {\bibinfo
  {title} {{Probing the Short-Distance Structure of the Quark-Gluon Plasma with
  Energy Correlators}},\ }\href
  {https://doi.org/10.1103/PhysRevLett.132.011901} {\bibfield  {journal}
  {\bibinfo  {journal} {Phys. Rev. Lett.}\ }\textbf {\bibinfo {volume} {132}},\
  \bibinfo {pages} {011901} (\bibinfo {year} {2024})},\ \Eprint
  {https://arxiv.org/abs/2310.01500} {arXiv:2310.01500 [hep-ph]} \BibitemShut
  {NoStop}%
\bibitem [{\citenamefont {Chekhovsky}\ \emph {et~al.}(2025)\citenamefont
  {Chekhovsky} \emph {et~al.}}]{CMS:2025ydi}%
  \BibitemOpen
  \bibfield  {author} {\bibinfo {author} {\bibfnamefont {V.}~\bibnamefont
  {Chekhovsky}} \emph {et~al.} (\bibinfo {collaboration} {CMS}),\ }\bibfield
  {title} {\bibinfo {title} {{Observation of nuclear modification of
  energy-energy correlators inside jets in heavy ion collisions}},\ }\href
  {https://doi.org/10.1016/j.physletb.2025.139556} {\bibfield  {journal}
  {\bibinfo  {journal} {Phys. Lett. B}\ }\textbf {\bibinfo {volume} {866}},\
  \bibinfo {pages} {139556} (\bibinfo {year} {2025})},\ \Eprint
  {https://arxiv.org/abs/2503.19993} {arXiv:2503.19993 [nucl-ex]} \BibitemShut
  {NoStop}%
\bibitem [{\citenamefont {Andres}\ \emph {et~al.}(2023)\citenamefont {Andres},
  \citenamefont {Dominguez}, \citenamefont {Kunnawalkam~Elayavalli},
  \citenamefont {Holguin}, \citenamefont {Marquet},\ and\ \citenamefont
  {Moult}}]{Andres:2022ovj}%
  \BibitemOpen
  \bibfield  {author} {\bibinfo {author} {\bibfnamefont {C.}~\bibnamefont
  {Andres}}, \bibinfo {author} {\bibfnamefont {F.}~\bibnamefont {Dominguez}},
  \bibinfo {author} {\bibfnamefont {R.}~\bibnamefont {Kunnawalkam~Elayavalli}},
  \bibinfo {author} {\bibfnamefont {J.}~\bibnamefont {Holguin}}, \bibinfo
  {author} {\bibfnamefont {C.}~\bibnamefont {Marquet}},\ and\ \bibinfo {author}
  {\bibfnamefont {I.}~\bibnamefont {Moult}},\ }\bibfield  {title} {\bibinfo
  {title} {{Resolving the Scales of the Quark-Gluon Plasma with Energy
  Correlators}},\ }\href {https://doi.org/10.1103/PhysRevLett.130.262301}
  {\bibfield  {journal} {\bibinfo  {journal} {Phys. Rev. Lett.}\ }\textbf
  {\bibinfo {volume} {130}},\ \bibinfo {pages} {262301} (\bibinfo {year}
  {2023})},\ \Eprint {https://arxiv.org/abs/2209.11236} {arXiv:2209.11236
  [hep-ph]} \BibitemShut {NoStop}%
\bibitem [{\citenamefont {Barata}\ \emph
  {et~al.}(2025{\natexlab{a}})\citenamefont {Barata}, \citenamefont {Brewer},
  \citenamefont {Lee},\ and\ \citenamefont {Silva}}]{Barata:2025uxp}%
  \BibitemOpen
  \bibfield  {author} {\bibinfo {author} {\bibfnamefont {J.}~\bibnamefont
  {Barata}}, \bibinfo {author} {\bibfnamefont {J.}~\bibnamefont {Brewer}},
  \bibinfo {author} {\bibfnamefont {K.}~\bibnamefont {Lee}},\ and\ \bibinfo
  {author} {\bibfnamefont {J.~M.}\ \bibnamefont {Silva}},\ }\bibfield  {title}
  {\bibinfo {title} {{Heavy Quark Pair Energy Correlators: From Profiling
  Partonic Splittings to Probing Heavy-Flavor Fragmentation}},\ }\href@noop {}
  {\  (\bibinfo {year} {2025}{\natexlab{a}})},\ \Eprint
  {https://arxiv.org/abs/2508.19404} {arXiv:2508.19404 [hep-ph]} \BibitemShut
  {NoStop}%
\bibitem [{\citenamefont {Apolin{\'a}rio}\ \emph {et~al.}(2025)\citenamefont
  {Apolin{\'a}rio}, \citenamefont {Kunnawalkam~Elayavalli}, \citenamefont
  {Madureira}, \citenamefont {Sheng}, \citenamefont {Wang},\ and\ \citenamefont
  {Yang}}]{Apolinario:2025vtx}%
  \BibitemOpen
  \bibfield  {author} {\bibinfo {author} {\bibfnamefont {L.}~\bibnamefont
  {Apolin{\'a}rio}}, \bibinfo {author} {\bibfnamefont {R.}~\bibnamefont
  {Kunnawalkam~Elayavalli}}, \bibinfo {author} {\bibfnamefont {N.~O.}\
  \bibnamefont {Madureira}}, \bibinfo {author} {\bibfnamefont {J.-X.}\
  \bibnamefont {Sheng}}, \bibinfo {author} {\bibfnamefont {X.-N.}\ \bibnamefont
  {Wang}},\ and\ \bibinfo {author} {\bibfnamefont {Z.}~\bibnamefont {Yang}},\
  }\bibfield  {title} {\bibinfo {title} {{Flavor dependence of energy-energy
  correlators}},\ }\href {https://doi.org/10.1103/4sfq-315y} {\bibfield
  {journal} {\bibinfo  {journal} {Phys. Rev. D}\ }\textbf {\bibinfo {volume}
  {112}},\ \bibinfo {pages} {054018} (\bibinfo {year} {2025})},\ \Eprint
  {https://arxiv.org/abs/2502.11406} {arXiv:2502.11406 [hep-ph]} \BibitemShut
  {NoStop}%
\bibitem [{\citenamefont {Xing}\ \emph {et~al.}(2025)\citenamefont {Xing},
  \citenamefont {Cao}, \citenamefont {Qin},\ and\ \citenamefont
  {Wang}}]{Xing:2024yrb}%
  \BibitemOpen
  \bibfield  {author} {\bibinfo {author} {\bibfnamefont {W.-J.}\ \bibnamefont
  {Xing}}, \bibinfo {author} {\bibfnamefont {S.}~\bibnamefont {Cao}}, \bibinfo
  {author} {\bibfnamefont {G.-Y.}\ \bibnamefont {Qin}},\ and\ \bibinfo {author}
  {\bibfnamefont {X.-N.}\ \bibnamefont {Wang}},\ }\bibfield  {title} {\bibinfo
  {title} {{Flavor Hierarchy of Jet Energy Correlators inside the Quark-Gluon
  Plasma}},\ }\href {https://doi.org/10.1103/PhysRevLett.134.052301} {\bibfield
   {journal} {\bibinfo  {journal} {Phys. Rev. Lett.}\ }\textbf {\bibinfo
  {volume} {134}},\ \bibinfo {pages} {052301} (\bibinfo {year} {2025})},\
  \Eprint {https://arxiv.org/abs/2409.12843} {arXiv:2409.12843 [hep-ph]}
  \BibitemShut {NoStop}%
\bibitem [{\citenamefont {Li}\ \emph {et~al.}(2021)\citenamefont {Li},
  \citenamefont {Makris},\ and\ \citenamefont {Vitev}}]{Li:2021txc}%
  \BibitemOpen
  \bibfield  {author} {\bibinfo {author} {\bibfnamefont {H.~T.}\ \bibnamefont
  {Li}}, \bibinfo {author} {\bibfnamefont {Y.}~\bibnamefont {Makris}},\ and\
  \bibinfo {author} {\bibfnamefont {I.}~\bibnamefont {Vitev}},\ }\bibfield
  {title} {\bibinfo {title} {{Energy-energy correlators in Deep Inelastic
  Scattering}},\ }\href {https://doi.org/10.1103/PhysRevD.103.094005}
  {\bibfield  {journal} {\bibinfo  {journal} {Phys. Rev. D}\ }\textbf {\bibinfo
  {volume} {103}},\ \bibinfo {pages} {094005} (\bibinfo {year} {2021})},\
  \Eprint {https://arxiv.org/abs/2102.05669} {arXiv:2102.05669 [hep-ph]}
  \BibitemShut {NoStop}%
\bibitem [{\citenamefont {Fu}\ \emph {et~al.}(2025)\citenamefont {Fu},
  \citenamefont {M{\"u}ller},\ and\ \citenamefont {Sirimanna}}]{Fu:2025gxu}%
  \BibitemOpen
  \bibfield  {author} {\bibinfo {author} {\bibfnamefont {Y.}~\bibnamefont
  {Fu}}, \bibinfo {author} {\bibfnamefont {B.}~\bibnamefont {M{\"u}ller}},\
  and\ \bibinfo {author} {\bibfnamefont {C.}~\bibnamefont {Sirimanna}},\
  }\bibfield  {title} {\bibinfo {title} {{Jet Energy-Energy Correlator in Cold
  QCD Matter}},\ }in\ \href@noop {} {\emph {\bibinfo {booktitle} {{31st
  International Conference on Ultra-relativistic Nucleus-Nucleus
  Collisions}}}}\ (\bibinfo {year} {2025})\ \Eprint
  {https://arxiv.org/abs/2510.08927} {arXiv:2510.08927 [hep-ph]} \BibitemShut
  {NoStop}%
\bibitem [{\citenamefont {Gonzalez}(2022)}]{Gonzalez:2022ulr}%
  \BibitemOpen
  \bibfield  {author} {\bibinfo {author} {\bibfnamefont {V.}~\bibnamefont
  {Gonzalez}} (\bibinfo {collaboration} {ALICE}),\ }\bibfield  {title}
  {\bibinfo {title} {{Characterizing system dynamics with two-particle
  transverse momentum correlations in pp collisions at $\sqrt{s} =
  7\;\text{TeV}$ and p{\textendash}Pb collisions at $\sqrt{s_{NN}} =
  5.02\;\text{TeV}$}},\ }\href {https://doi.org/10.22323/1.414.0930} {\bibfield
   {journal} {\bibinfo  {journal} {PoS}\ }\textbf {\bibinfo {volume}
  {ICHEP2022}},\ \bibinfo {pages} {930} (\bibinfo {year} {2022})},\ \Eprint
  {https://arxiv.org/abs/2211.10467} {arXiv:2211.10467 [nucl-ex]} \BibitemShut
  {NoStop}%
\bibitem [{\citenamefont {Barata}\ \emph
  {et~al.}(2025{\natexlab{b}})\citenamefont {Barata}, \citenamefont {Kang},
  \citenamefont {Mayo~L{\'o}pez},\ and\ \citenamefont
  {Penttala}}]{Barata:2024wsu}%
  \BibitemOpen
  \bibfield  {author} {\bibinfo {author} {\bibfnamefont {J.}~\bibnamefont
  {Barata}}, \bibinfo {author} {\bibfnamefont {Z.-B.}\ \bibnamefont {Kang}},
  \bibinfo {author} {\bibfnamefont {X.}~\bibnamefont {Mayo~L{\'o}pez}},\ and\
  \bibinfo {author} {\bibfnamefont {J.}~\bibnamefont {Penttala}},\ }\bibfield
  {title} {\bibinfo {title} {{Energy-Energy Correlator for Jet Production in pp
  and pA Collisions}},\ }\href {https://doi.org/10.1103/96xh-bd1w} {\bibfield
  {journal} {\bibinfo  {journal} {Phys. Rev. Lett.}\ }\textbf {\bibinfo
  {volume} {134}},\ \bibinfo {pages} {251903} (\bibinfo {year}
  {2025}{\natexlab{b}})},\ \Eprint {https://arxiv.org/abs/2411.11782}
  {arXiv:2411.11782 [hep-ph]} \BibitemShut {NoStop}%
\bibitem [{\citenamefont {Bossi}\ \emph {et~al.}(2025)\citenamefont {Bossi}
  \emph {et~al.}}]{Electron-PositronAlliance:2025fhk}%
  \BibitemOpen
  \bibfield  {author} {\bibinfo {author} {\bibfnamefont {H.}~\bibnamefont
  {Bossi}} \emph {et~al.} (\bibinfo {collaboration} {Electron-Positron
  Alliance}),\ }\bibfield  {title} {\bibinfo {title} {{Energy Correlators from
  Partons to Hadrons: Unveiling the Dynamics of the Strong Interactions with
  Archival ALEPH Data}},\ }\href@noop {} {\  (\bibinfo {year} {2025})},\
  \Eprint {https://arxiv.org/abs/2511.00149} {arXiv:2511.00149 [hep-ph]}
  \BibitemShut {NoStop}%
\bibitem [{\citenamefont {Cacciari}\ \emph {et~al.}(2008)\citenamefont
  {Cacciari}, \citenamefont {Salam},\ and\ \citenamefont
  {Soyez}}]{Cacciari:2008gp}%
  \BibitemOpen
  \bibfield  {author} {\bibinfo {author} {\bibfnamefont {M.}~\bibnamefont
  {Cacciari}}, \bibinfo {author} {\bibfnamefont {G.~P.}\ \bibnamefont
  {Salam}},\ and\ \bibinfo {author} {\bibfnamefont {G.}~\bibnamefont {Soyez}},\
  }\bibfield  {title} {\bibinfo {title} {{The anti-$k_t$ jet clustering
  algorithm}},\ }\href {https://doi.org/10.1088/1126-6708/2008/04/063}
  {\bibfield  {journal} {\bibinfo  {journal} {JHEP}\ }\textbf {\bibinfo
  {volume} {04}},\ \bibinfo {pages} {063}},\ \Eprint
  {https://arxiv.org/abs/0802.1189} {arXiv:0802.1189 [hep-ph]} \BibitemShut
  {NoStop}%
\bibitem [{\citenamefont {Aad}\ \emph {et~al.}(2019)\citenamefont {Aad} \emph
  {et~al.}}]{ATLAS:2019ocl}%
  \BibitemOpen
  \bibfield  {author} {\bibinfo {author} {\bibfnamefont {G.}~\bibnamefont
  {Aad}} \emph {et~al.} (\bibinfo {collaboration} {ATLAS}),\ }\bibfield
  {title} {\bibinfo {title} {{Measurement of distributions sensitive to the
  underlying event in inclusive $Z$-boson production in pp collisions at
  $\sqrt{s} = 13$ TeV with the ATLAS detector}},\ }\href
  {https://doi.org/10.1140/epjc/s10052-019-7162-0} {\bibfield  {journal}
  {\bibinfo  {journal} {Eur. Phys. J. C}\ }\textbf {\bibinfo {volume} {79}},\
  \bibinfo {pages} {666} (\bibinfo {year} {2019})},\ \Eprint
  {https://arxiv.org/abs/1905.09752} {arXiv:1905.09752 [hep-ex]} \BibitemShut
  {NoStop}%
\bibitem [{\citenamefont {Abelev}\ \emph {et~al.}(2012)\citenamefont {Abelev}
  \emph {et~al.}}]{ALICE:2011ac}%
  \BibitemOpen
  \bibfield  {author} {\bibinfo {author} {\bibfnamefont {B.}~\bibnamefont
  {Abelev}} \emph {et~al.} (\bibinfo {collaboration} {ALICE}),\ }\bibfield
  {title} {\bibinfo {title} {{Underlying Event measurements in $pp$ collisions
  at $\sqrt{s}=0.9$ and 7 TeV with the ALICE experiment at the LHC}},\ }\href
  {https://doi.org/10.1007/JHEP07(2012)116} {\bibfield  {journal} {\bibinfo
  {journal} {JHEP}\ }\textbf {\bibinfo {volume} {07}},\ \bibinfo {pages}
  {116}},\ \Eprint {https://arxiv.org/abs/1112.2082} {arXiv:1112.2082 [hep-ex]}
  \BibitemShut {NoStop}%
\bibitem [{\citenamefont {Peng}\ \emph {et~al.}(2025)\citenamefont {Peng},
  \citenamefont {Wu}, \citenamefont {Peng}, \citenamefont {Yin},\ and\
  \citenamefont {Zheng}}]{Peng:2025mpf}%
  \BibitemOpen
  \bibfield  {author} {\bibinfo {author} {\bibfnamefont {Y.}~\bibnamefont
  {Peng}}, \bibinfo {author} {\bibfnamefont {Y.}~\bibnamefont {Wu}}, \bibinfo
  {author} {\bibfnamefont {X.}~\bibnamefont {Peng}}, \bibinfo {author}
  {\bibfnamefont {Z.}~\bibnamefont {Yin}},\ and\ \bibinfo {author}
  {\bibfnamefont {L.}~\bibnamefont {Zheng}},\ }\bibfield  {title} {\bibinfo
  {title} {{Investigating jet-induced identified hadron production from the
  relative transverse activity classifier in pp collisions at the LHC}},\
  }\href {https://doi.org/10.1103/bm5b-829j} {\bibfield  {journal} {\bibinfo
  {journal} {Phys. Rev. D}\ }\textbf {\bibinfo {volume} {112}},\ \bibinfo
  {pages} {114010} (\bibinfo {year} {2025})},\ \Eprint
  {https://arxiv.org/abs/2507.23306} {arXiv:2507.23306 [hep-ph]} \BibitemShut
  {NoStop}%
\bibitem [{\citenamefont {Acharya}\ \emph {et~al.}(2020)\citenamefont {Acharya}
  \emph {et~al.}}]{ALICE:2019mmy}%
  \BibitemOpen
  \bibfield  {author} {\bibinfo {author} {\bibfnamefont {S.}~\bibnamefont
  {Acharya}} \emph {et~al.} (\bibinfo {collaboration} {ALICE}),\ }\bibfield
  {title} {\bibinfo {title} {{Underlying Event properties in pp collisions at
  $\sqrt{s}$ = 13 TeV}},\ }\href {https://doi.org/10.1007/JHEP04(2020)192}
  {\bibfield  {journal} {\bibinfo  {journal} {JHEP}\ }\textbf {\bibinfo
  {volume} {04}},\ \bibinfo {pages} {192}},\ \Eprint
  {https://arxiv.org/abs/1910.14400} {arXiv:1910.14400 [nucl-ex]} \BibitemShut
  {NoStop}%
\bibitem [{\citenamefont {Bencedi}\ \emph {et~al.}(2021)\citenamefont
  {Bencedi}, \citenamefont {Ortiz},\ and\ \citenamefont
  {Paz}}]{Bencedi:2021tst}%
  \BibitemOpen
  \bibfield  {author} {\bibinfo {author} {\bibfnamefont {G.}~\bibnamefont
  {Bencedi}}, \bibinfo {author} {\bibfnamefont {A.}~\bibnamefont {Ortiz}},\
  and\ \bibinfo {author} {\bibfnamefont {A.}~\bibnamefont {Paz}},\ }\bibfield
  {title} {\bibinfo {title} {{Disentangling the hard gluon bremsstrahlung
  effects from the relative transverse activity classifier in pp collisions}},\
  }\href {https://doi.org/10.1103/PhysRevD.104.016017} {\bibfield  {journal}
  {\bibinfo  {journal} {Phys. Rev. D}\ }\textbf {\bibinfo {volume} {104}},\
  \bibinfo {pages} {016017} (\bibinfo {year} {2021})},\ \Eprint
  {https://arxiv.org/abs/2105.04838} {arXiv:2105.04838 [hep-ph]} \BibitemShut
  {NoStop}%
\bibitem [{\citenamefont {Acosta}\ \emph {et~al.}(2004)\citenamefont {Acosta}
  \emph {et~al.}}]{CDF:2004jod}%
  \BibitemOpen
  \bibfield  {author} {\bibinfo {author} {\bibfnamefont {D.}~\bibnamefont
  {Acosta}} \emph {et~al.} (\bibinfo {collaboration} {CDF}),\ }\bibfield
  {title} {\bibinfo {title} {{The underlying event in hard interactions at the
  Tevatron $\bar{p}p$ collider}},\ }\href
  {https://doi.org/10.1103/PhysRevD.70.072002} {\bibfield  {journal} {\bibinfo
  {journal} {Phys. Rev. D}\ }\textbf {\bibinfo {volume} {70}},\ \bibinfo
  {pages} {072002} (\bibinfo {year} {2004})},\ \Eprint
  {https://arxiv.org/abs/hep-ex/0404004} {arXiv:hep-ex/0404004} \BibitemShut
  {NoStop}%
\bibitem [{\citenamefont {Verma}\ \emph {et~al.}(2025)\citenamefont {Verma},
  \citenamefont {Saini}, \citenamefont {Nandi},\ and\ \citenamefont
  {Dash}}]{Verma:2024fry}%
  \BibitemOpen
  \bibfield  {author} {\bibinfo {author} {\bibfnamefont {R.}~\bibnamefont
  {Verma}}, \bibinfo {author} {\bibfnamefont {V.}~\bibnamefont {Saini}},
  \bibinfo {author} {\bibfnamefont {B.~K.}\ \bibnamefont {Nandi}},\ and\
  \bibinfo {author} {\bibfnamefont {S.}~\bibnamefont {Dash}},\ }\bibfield
  {title} {\bibinfo {title} {{Study of identified particle production as a
  function of transverse event activity classifier, $S_{T}$ in
  p{\ensuremath{-}}p collisions at LHC energies}},\ }\href
  {https://doi.org/10.1140/epjp/s13360-024-05951-0} {\bibfield  {journal}
  {\bibinfo  {journal} {Eur. Phys. J. Plus}\ }\textbf {\bibinfo {volume}
  {140}},\ \bibinfo {pages} {62} (\bibinfo {year} {2025})},\ \Eprint
  {https://arxiv.org/abs/2403.01224} {arXiv:2403.01224 [hep-ph]} \BibitemShut
  {NoStop}%
\bibitem [{\citenamefont {Field}(2001)}]{Field:2001aok}%
  \BibitemOpen
  \bibfield  {author} {\bibinfo {author} {\bibfnamefont {R.~D.}\ \bibnamefont
  {Field}} (\bibinfo {collaboration} {CDF}),\ }\bibfield  {title} {\bibinfo
  {title} {{The Underlying Event in Hard Scattering Processes}},\ }\href@noop
  {} {\bibfield  {journal} {\bibinfo  {journal} {eConf}\ }\textbf {\bibinfo
  {volume} {C010630}},\ \bibinfo {pages} {P501} (\bibinfo {year} {2001})},\
  \Eprint {https://arxiv.org/abs/hep-ph/0201192} {arXiv:hep-ph/0201192}
  \BibitemShut {NoStop}%
\bibitem [{\citenamefont {Bierlich}\ \emph {et~al.}(2022)\citenamefont
  {Bierlich} \emph {et~al.}}]{Bierlich:2022pfr}%
  \BibitemOpen
  \bibfield  {author} {\bibinfo {author} {\bibfnamefont {C.}~\bibnamefont
  {Bierlich}} \emph {et~al.},\ }\bibfield  {title} {\bibinfo {title} {{A
  comprehensive guide to the physics and usage of PYTHIA 8.3}},\ }\href
  {https://doi.org/10.21468/SciPostPhysCodeb.8} {\bibfield  {journal} {\bibinfo
   {journal} {SciPost Phys. Codeb.}\ }\textbf {\bibinfo {volume} {2022}},\
  \bibinfo {pages} {8} (\bibinfo {year} {2022})},\ \Eprint
  {https://arxiv.org/abs/2203.11601} {arXiv:2203.11601 [hep-ph]} \BibitemShut
  {NoStop}%
\bibitem [{\citenamefont {Cacciari}\ \emph {et~al.}(2012)\citenamefont
  {Cacciari}, \citenamefont {Salam},\ and\ \citenamefont
  {Soyez}}]{Cacciari:2011ma}%
  \BibitemOpen
  \bibfield  {author} {\bibinfo {author} {\bibfnamefont {M.}~\bibnamefont
  {Cacciari}}, \bibinfo {author} {\bibfnamefont {G.~P.}\ \bibnamefont
  {Salam}},\ and\ \bibinfo {author} {\bibfnamefont {G.}~\bibnamefont {Soyez}},\
  }\bibfield  {title} {\bibinfo {title} {{FastJet User Manual}},\ }\href
  {https://doi.org/10.1140/epjc/s10052-012-1896-2} {\bibfield  {journal}
  {\bibinfo  {journal} {Eur. Phys. J. C}\ }\textbf {\bibinfo {volume} {72}},\
  \bibinfo {pages} {1896} (\bibinfo {year} {2012})},\ \Eprint
  {https://arxiv.org/abs/1111.6097} {arXiv:1111.6097 [hep-ph]} \BibitemShut
  {NoStop}%
\bibitem [{\citenamefont {Larkoski}\ \emph {et~al.}(2014)\citenamefont
  {Larkoski}, \citenamefont {Neill},\ and\ \citenamefont
  {Thaler}}]{Larkoski:2014uqa}%
  \BibitemOpen
  \bibfield  {author} {\bibinfo {author} {\bibfnamefont {A.~J.}\ \bibnamefont
  {Larkoski}}, \bibinfo {author} {\bibfnamefont {D.}~\bibnamefont {Neill}},\
  and\ \bibinfo {author} {\bibfnamefont {J.}~\bibnamefont {Thaler}},\
  }\bibfield  {title} {\bibinfo {title} {{Jet Shapes with the Broadening
  Axis}},\ }\href {https://doi.org/10.1007/JHEP04(2014)017} {\bibfield
  {journal} {\bibinfo  {journal} {JHEP}\ }\textbf {\bibinfo {volume} {04}},\
  \bibinfo {pages} {017}},\ \Eprint {https://arxiv.org/abs/1401.2158}
  {arXiv:1401.2158 [hep-ph]} \BibitemShut {NoStop}%
\bibitem [{\citenamefont {Cacciari}\ \emph {et~al.}(2015)\citenamefont
  {Cacciari}, \citenamefont {Salam},\ and\ \citenamefont
  {Soyez}}]{Cacciari:2014gra}%
  \BibitemOpen
  \bibfield  {author} {\bibinfo {author} {\bibfnamefont {M.}~\bibnamefont
  {Cacciari}}, \bibinfo {author} {\bibfnamefont {G.~P.}\ \bibnamefont
  {Salam}},\ and\ \bibinfo {author} {\bibfnamefont {G.}~\bibnamefont {Soyez}},\
  }\bibfield  {title} {\bibinfo {title} {{SoftKiller, a particle-level pileup
  removal method}},\ }\href {https://doi.org/10.1140/epjc/s10052-015-3267-2}
  {\bibfield  {journal} {\bibinfo  {journal} {Eur. Phys. J. C}\ }\textbf
  {\bibinfo {volume} {75}},\ \bibinfo {pages} {59} (\bibinfo {year} {2015})},\
  \Eprint {https://arxiv.org/abs/1407.0408} {arXiv:1407.0408 [hep-ph]}
  \BibitemShut {NoStop}%
\bibitem [{\citenamefont {Cal}\ \emph {et~al.}(2020)\citenamefont {Cal},
  \citenamefont {Neill}, \citenamefont {Ringer},\ and\ \citenamefont
  {Waalewijn}}]{Cal:2019gxa}%
  \BibitemOpen
  \bibfield  {author} {\bibinfo {author} {\bibfnamefont {P.}~\bibnamefont
  {Cal}}, \bibinfo {author} {\bibfnamefont {D.}~\bibnamefont {Neill}}, \bibinfo
  {author} {\bibfnamefont {F.}~\bibnamefont {Ringer}},\ and\ \bibinfo {author}
  {\bibfnamefont {W.~J.}\ \bibnamefont {Waalewijn}},\ }\bibfield  {title}
  {\bibinfo {title} {{Calculating the angle between jet axes}},\ }\href
  {https://doi.org/10.1007/JHEP04(2020)211} {\bibfield  {journal} {\bibinfo
  {journal} {JHEP}\ }\textbf {\bibinfo {volume} {04}},\ \bibinfo {pages}
  {211}},\ \Eprint {https://arxiv.org/abs/1911.06840} {arXiv:1911.06840
  [hep-ph]} \BibitemShut {NoStop}%
\bibitem [{\citenamefont {Hayrapetyan}\ \emph {et~al.}(2024)\citenamefont
  {Hayrapetyan} \emph {et~al.}}]{CMS:2024mlf}%
  \BibitemOpen
  \bibfield  {author} {\bibinfo {author} {\bibfnamefont {A.}~\bibnamefont
  {Hayrapetyan}} \emph {et~al.} (\bibinfo {collaboration} {CMS}),\ }\bibfield
  {title} {\bibinfo {title} {{Measurement of Energy Correlators inside Jets and
  Determination of the Strong Coupling {\ensuremath{\alpha}}S(mZ)}},\ }\href
  {https://doi.org/10.1103/PhysRevLett.133.071903} {\bibfield  {journal}
  {\bibinfo  {journal} {Phys. Rev. Lett.}\ }\textbf {\bibinfo {volume} {133}},\
  \bibinfo {pages} {071903} (\bibinfo {year} {2024})},\ \Eprint
  {https://arxiv.org/abs/2402.13864} {arXiv:2402.13864 [hep-ex]} \BibitemShut
  {NoStop}%
\bibitem [{\citenamefont {Chen}\ \emph {et~al.}(2026)\citenamefont {Chen},
  \citenamefont {Xu}, \citenamefont {Shen}, \citenamefont {Dai}, \citenamefont
  {Zhang},\ and\ \citenamefont {Wang}}]{Chen:2024quk}%
  \BibitemOpen
  \bibfield  {author} {\bibinfo {author} {\bibfnamefont {S.-Y.}\ \bibnamefont
  {Chen}}, \bibinfo {author} {\bibfnamefont {Z.-X.}\ \bibnamefont {Xu}},
  \bibinfo {author} {\bibfnamefont {K.-M.}\ \bibnamefont {Shen}}, \bibinfo
  {author} {\bibfnamefont {W.}~\bibnamefont {Dai}}, \bibinfo {author}
  {\bibfnamefont {B.-W.}\ \bibnamefont {Zhang}},\ and\ \bibinfo {author}
  {\bibfnamefont {E.}~\bibnamefont {Wang}},\ }\bibfield  {title} {\bibinfo
  {title} {{Study of the EEC discrimination power on quark and gluon jet
  quenching effects in heavy-ion collisions at${\sqrt{s}=}$ 5.02 TeV*}},\
  }\href {https://doi.org/10.1088/1674-1137/ae056c} {\bibfield  {journal}
  {\bibinfo  {journal} {Chin. Phys.}\ }\textbf {\bibinfo {volume} {50}},\
  \bibinfo {pages} {023114} (\bibinfo {year} {2026})},\ \Eprint
  {https://arxiv.org/abs/2409.13996} {arXiv:2409.13996 [nucl-th]} \BibitemShut
  {NoStop}%
\bibitem [{\citenamefont {Forte}(2025)}]{Forte:2025twm}%
  \BibitemOpen
  \bibfield  {author} {\bibinfo {author} {\bibfnamefont {S.}~\bibnamefont
  {Forte}},\ }\bibfield  {title} {\bibinfo {title} {{Asymptotic Freedom in
  Parton Language: the Birth of Perturbative QCD}}\ }(\bibinfo {year} {2025})\
  \Eprint {https://arxiv.org/abs/2501.13158} {arXiv:2501.13158 [hep-ph]}
  \BibitemShut {NoStop}%
\bibitem [{\citenamefont {Dokshitzer}\ \emph {et~al.}(1991)\citenamefont
  {Dokshitzer}, \citenamefont {Khoze},\ and\ \citenamefont
  {Troian}}]{Dokshitzer:1991fd}%
  \BibitemOpen
  \bibfield  {author} {\bibinfo {author} {\bibfnamefont {Y.~L.}\ \bibnamefont
  {Dokshitzer}}, \bibinfo {author} {\bibfnamefont {V.~A.}\ \bibnamefont
  {Khoze}},\ and\ \bibinfo {author} {\bibfnamefont {S.~I.}\ \bibnamefont
  {Troian}},\ }\bibfield  {title} {\bibinfo {title} {{On specific QCD
  properties of heavy quark fragmentation ('dead cone')}},\ }\href
  {https://doi.org/10.1088/0954-3899/17/10/023} {\bibfield  {journal} {\bibinfo
   {journal} {J. Phys. G}\ }\textbf {\bibinfo {volume} {17}},\ \bibinfo {pages}
  {1602} (\bibinfo {year} {1991})}\BibitemShut {NoStop}%
\bibitem [{\citenamefont {Acharya}\ \emph {et~al.}(2022)\citenamefont {Acharya}
  \emph {et~al.}}]{ALICE:2021aqk}%
  \BibitemOpen
  \bibfield  {author} {\bibinfo {author} {\bibfnamefont {S.}~\bibnamefont
  {Acharya}} \emph {et~al.} (\bibinfo {collaboration} {ALICE}),\ }\bibfield
  {title} {\bibinfo {title} {{Direct observation of the dead-cone effect in
  quantum chromodynamics}},\ }\href
  {https://doi.org/10.1038/s41586-022-04572-w} {\bibfield  {journal} {\bibinfo
  {journal} {Nature}\ }\textbf {\bibinfo {volume} {605}},\ \bibinfo {pages}
  {440} (\bibinfo {year} {2022})},\ \bibinfo {note} {[Erratum: Nature 607, E22
  (2022)]},\ \Eprint {https://arxiv.org/abs/2106.05713} {arXiv:2106.05713
  [nucl-ex]} \BibitemShut {NoStop}%
\bibitem [{\citenamefont {Craft}\ \emph {et~al.}(2022)\citenamefont {Craft},
  \citenamefont {Lee}, \citenamefont {Me{\c{c}}aj},\ and\ \citenamefont
  {Moult}}]{Craft:2022kdo}%
  \BibitemOpen
  \bibfield  {author} {\bibinfo {author} {\bibfnamefont {E.}~\bibnamefont
  {Craft}}, \bibinfo {author} {\bibfnamefont {K.}~\bibnamefont {Lee}}, \bibinfo
  {author} {\bibfnamefont {B.}~\bibnamefont {Me{\c{c}}aj}},\ and\ \bibinfo
  {author} {\bibfnamefont {I.}~\bibnamefont {Moult}},\ }\bibfield  {title}
  {\bibinfo {title} {{Beautiful and Charming Energy Correlators}},\ }\href@noop
  {} {\  (\bibinfo {year} {2022})},\ \Eprint {https://arxiv.org/abs/2210.09311}
  {arXiv:2210.09311 [hep-ph]} \BibitemShut {NoStop}%
\bibitem [{\citenamefont {Acharya}\ \emph {et~al.}(2025)\citenamefont {Acharya}
  \emph {et~al.}}]{ALICE:2025igw}%
  \BibitemOpen
  \bibfield  {author} {\bibinfo {author} {\bibfnamefont {S.}~\bibnamefont
  {Acharya}} \emph {et~al.} (\bibinfo {collaboration} {ALICE}),\ }\bibfield
  {title} {\bibinfo {title} {{Energy-energy correlators in charm-tagged jets in
  proton-proton collisions at $\mathbf{\sqrt{s} = 13}$ TeV}},\ }\href@noop {}
  {\  (\bibinfo {year} {2025})},\ \Eprint {https://arxiv.org/abs/2504.03431}
  {arXiv:2504.03431 [hep-ex]} \BibitemShut {NoStop}%
\end{thebibliography}%

\end{document}